\documentclass{aa}  
\usepackage{graphicx}
\usepackage{txfonts}
\usepackage[draft]{hyperref}
\usepackage[usenames, dvipsnames]{color}
\usepackage{enumitem}
\usepackage{ulem}
\usepackage{lscape}
\usepackage{soul}
\begin{document} 

\renewcommand{\arraystretch}{1.5}
   \title{The Dispersion Measure Contributions of the Cosmic Web}

   \titlerunning{DMs of the Cosmic Web} 
   \authorrunning{Walker et al.}
   
   \author{Charles R. H. Walker\inst{1},
          Laura G. Spitler\inst{1},
          Yin-Zhe Ma\inst{2},
          Cheng Cheng\inst{3},
          M. Celeste Artale\inst{4,5,6},
          \and
          Cameron Hummels\inst{7}
          }

   \institute{Max-Planck-Institute f{\"u}r Radioastronomie, Auf dem H{\"u}gel 69, D-53121, Bonn, Germany \\
   \url{Emails: cwalker@mpifr-bonn.mpg.de}, \url{lspitler@mpifr-bonn.mpg.de}
         \and
         Department of Physics, Stellenbosch University, Matieland 7602, South Africa.
         \url{Email: mayinzhe@sun.ac.za}
         \and
         Xinjiang Astronomical Observatory, Chinese Academy of Sciences, Urumqi 830011, China
          \and
            Instituto de Astrofisica, Facultad de Ciencias Exactas, Universidad Andres Bello, Fernandez Concha 700, Santiago, Chile \and
            Physics and Astronomy Department Galileo Galilei, University of Padova, Vicolo dell'Osservatorio 3, I--35122, Padova, Italy 
            \and
           INFN--Padova, Via Marzolo 8, I--35131 Padova, Italy
          \and
            {TAPIR, California Institute of Technology, Pasadena, CA 91125, USA}
             }

   \date{\today}

 
  \abstract
   {The large-scale distribution of baryons is sensitive to gravitational collapse, mergers, and galactic feedback processes. Known colloquially as the Cosmic Web, its large-scale structure (LSS) can be classified as halos, filaments, and voids. Fast Radio Bursts (FRBs) are extragalactic transient radio sources that undergo dispersion along their propagation paths. These systems provide insight into ionised matter along their sightlines by virtue of their dispersion measures (DMs), and have been investigated as probes of the LSS baryon fraction, the diffuse baryon distribution, and of cosmological parameters. Such efforts are highly complementary to the study of intergalactic medium (IGM) through X-ray observations, the Sunyaev-Zeldovich effect, and galaxy populations.}
   {We use the cosmological simulation {\tt IllustrisTNG} to study FRB DMs accumulated while traversing different types of LSS.}
   {We combine methods for deriving electron density, classifying LSS, and tracing FRB sightlines through {\tt TNG300-1}. We identify halos, filaments, voids, and collapsed structures along randomly selected sightlines, and calculate their DM contributions.}
   {We present comprehensive analysis of the redshift-evolving cosmological DM components of the Cosmic Web. We find that the filamentary contribution to DM dominates, increasing from $\sim71\%$ to $\sim80\%$ on average for FRBs originating at $z=0.1$ vs $z=5$, while the halo contribution falls, and the void contribution remains consistent to within $\sim1\%$. The majority of DM variance between sightlines originates from halo and filamentary environments, potentially making void-only sightlines more precise probes of cosmological parameters. We find that, on average, an FRB originating at $z=1$ will intersect $\sim1.8$ foreground collapsed structures of any mass, increasing to $\sim12.4$ structures for an FRB originating at $z=5$. The measured impact parameters between our sightlines and {\tt TNG} structures of any mass appear consistent with those reported for likely galaxy-intersecting FRBs. However, we measure lower average accumulated DMs from these structures than the $\sim90\;{\rm pc\;cm^{-3}}$ DM excesses reported for these literature FRBs, indicating some of this DM may arise beyond the structures themselves.}
   {}


   \maketitle
%

\section{Introduction}\label{sect:1}

Since their discovery~\citep{lori07a}, the microsecond to millisecond-duration luminous signals known as fast radio bursts (FRBs) are being increasingly detected\footnote{For various FRB catalogues, see: \url{www.frbcat.org}~\citep{petr16a}, \url{https://cdsarc.cds.unistra.fr/viz-bin/cat/J/ApJS/257/59} \citep{CHIME21a}, and \url{https://www.wis-tns.org/} \citep{yaro20a}.}, and their natures heavily investigated~\citep{zhan18a,plat19a}. Though the catalogue size of localised sources expands \citep{hein20a}, the underlying progenitor(s) of these radio transients is not yet unambiguously determined \citep{plat19a}. One uncontroversial observation, however, is that most FRBs traverse extragalactic distances\footnote{Just one FRB-like source has thus far been associated with a Galactic counterpart~\citep{boch20a}.}, accumulating large dispersion measures (DMs), which can be measured from the frequency-dependent delay in their arrival times ($\Delta t\sim {\rm DM}/\nu^{2}$). DM is considered an acceptable proxy for the integrated electron density to the distant source of emission along the line-of-sight~\citep{kulk20a} 
\begin{eqnarray}
{\rm DM}=\int^{z}_{0} \frac{n_{\rm e}(z)}{(1+z)}\,{\rm d}l,
    \label{eq:DM_def}
\end{eqnarray}
where $n_{\rm e}$ is the physical number density of free electrons at redshift $z$, and ${\rm d}l$ is the proper distance increment \citep{macq20a}. Because of this definition, the total observed DM for an FRB can be decomposed into three parts \citep{walt18a}
\begin{eqnarray}
{\rm DM}_{\rm obs}={\rm DM}_{\rm MW}+{\rm DM}_{\rm cos}+\frac{{\rm DM}_{\rm host}}{1+z}, \label{eq:DM_obs}
\end{eqnarray}
where ${\rm DM}_{\rm MW}$, ${\rm DM}_{\rm cos}$ and ${\rm DM}_{\rm host}$ are the DM contributions from our Milky Way, the cosmological ionised medium and the FRB host galaxy respectively.

Work is underway to constrain ${\rm DM}_{\rm MW}$, which is direction-dependent, and encompasses the contributions of both the Milky Way's disk and its halo \citep{cook23a}. Large values of ${\rm DM}_{\rm host}$ have been predicted \citep{walk20a} and observed \citep{spit14a,niu22a} for some FRBs, but ${\rm DM}_{\rm host}$ is supressed by a factor of $(1+z)^{-1}$ for distant galaxies due to cosmological redshift and time dilation effects~\citep{Yang2016}. Therefore a major component of DM for a source at a large cosmological distance will often be ${\rm DM}_{\rm cos}$, which is attributed to a combination of the ionised matter in diffuse cosmological structures (also denoted ${\rm DM}_{\rm IGM}$), and the cirumgalactic media (CGM) of any foreground galaxies (often denoted ${\rm DM}_{\rm halos}$) along the line-of-sight~\citep{proc19a}. In general, ${\rm DM}_{\rm cos}$ can be written as 

\begin{eqnarray}
{\rm DM}_{\rm cos}(z) = c\int^{z_{\rm s}}_{0} \frac{n_{\rm e}(z){\rm d}z}{(1+z)^2{H(z)}},
\label{DMIGMeq}
\end{eqnarray}
with speed of light $c$ and Hubble parameter for a spacially-flat universe 
\begin{eqnarray}
H(z) & \equiv & H_{0}E(z) \nonumber \\
&=& H_{0}\sqrt{\Omega_{\rm m}(1+z)^2+\Omega_\Lambda},
\label{e_eq}
\end{eqnarray}
where $H_{0}$ is the Hubble constant. In Eq.\ (\ref{e_eq}), $\Omega_{\rm m}$ and $\Omega_{\Lambda}$ are the fractional densities of matter and dark energy respectively. From Eq.\ (\ref{DMIGMeq}), one can see that $n_{\rm e}$ depends on the ionisation state of the baryons along the sightline~\citep{inou04a}. Even for two FRBs originating at the same redshift, the different collapsed structures along their propagation paths can cause different values of ${\rm DM}_{\rm cos}$. However the average $\langle {\rm DM} \rangle$ for many FRBs on different sightlines has a predictable relationship with redshift, and \citet{macq20a} have used five localised FRBs to directly measure the baryon density of the Universe, finding a result consistent with measurements from the cosmic microwave background \citep{Planck_parameters} and Big-Bang nucleosynthesis \citep{Cooke2018}.

Although \citet{macq20a} determined the total baryon density, the detailed location of these baryons in the Cosmic Web are not well known. Due to  galactic feedback processes \citep{cen06a,breg07a}, a significant amount of baryonic matter is anticipated to reside in a so-called warm-hot intergalactic medium (hereafter WHIM) with temperatures $\sim10^{5}$-$10^{7}$\,Kelvin, which can be diffuse around halos, filaments and even in voids~\citep{haid16a,marti19a}. In recent years, methods leveraging the Sunyaev-Zeldovich effect \citep{Ma2015,Hojjati2015,Hojjati2017,Tanimura17,Ma2017,tani19a,Ma2021} and Oxygen absorption lines (OVII;~\citealt{Nicastro2018}) have been developed to probe the WHIM. As well as counting all electrons along FRB sightlines, FRB DMs may prove complementary to these distribution-sensitive probes (see, e.g., \citealt{inou04a,akah16a}). \citet{mcqu14a} showed that the distribution of ${\rm DM}_{\rm cos}$ is sensitive to the location of baryons within, or beyond galactic halos. \citet{walt19a} showed that with $10^{2}-10^{3}$ FRBs, the diffuse baryon fraction could be determined with a few percent error. \citet{lee21a} proposed to use FRBs, together with foreground matter distribution mapping and Bayesian reconstruction techniques, to constrain the baryon fractions contained in the CGM of galactic halos and other cosmological structures (see also \citealt{walk20a}).

In this work we therefore deconstruct the contributions to FRB DMs from different types of cosmological large-scale structure using the results of the numerical simulation suite {\tt IllustrisTNG}. {\tt IllustrisTNG} has already proven a valuable tool for studying the Cosmic Web's formation \citep{marti19a}, its influence on galaxies \citep{donn22a,mala22a}, and the portion of its matter which remains difficult to detect observationally \citep{pari22a}. We will thus use {\tt IllustrisTNG} to analyse the contributions to DM from halos, filaments and voids, and provide comprehensive empirical distributions of these as a function of redshift. In Sect.\ \ref{sect:2}, we introduce {\tt TNG300-1}, our simulation run of choice, and discuss the process of obtaining electron densities from it. In Sect.\ \ref{sect:3}, we focus on the significance of LSS, and define our criteria for classifying it within the simulation. In Sect.\ \ref{sect:4}, we discuss our method to calculate FRB DMs given the sightlines. We review our results in Sect.\ \ref{sect:5}, and discuss their implications in the context of other studies in Sect.\ \ref{sect:6}. We conclude in Sect.\ \ref{sect:7}.

To maintain consistency with {\tt IllustrisTNG}~\citep{nels19a}, unless otherwise stated we adopt the {\it Planck}-2015 cosmological parameters~\citep{plan15b}: fractional baryon density $\Omega_{\rm b}=0.0486$, fractional matter density $\Omega_{\rm m}=0.3089$, fractional dark energy density $\Omega_{\Lambda}=0.6911$, spectral index $n_{\rm s} = 0.9667$, {\it rms} fluctuation amplitude $\sigma_{8}=0.8159$, and reduced Hubble constant $h=0.6774$.

\section{Simulations}\label{sect:2}

\begin{table}
\caption[]{The `full' {\tt TNG300-1} snapshots used in this analysis.}
\label{table:snaps}
$$
\begin{array}{p{0.5\linewidth}l}
\hline
\noalign{\smallskip}
Snapshot Number & {\rm Redshift}\\
\noalign{\smallskip}
\hline
99 & 0.00\\
91 & 0.10\\
84 & 0.20\\
78 & 0.30\\
72 & 0.40\\
67 & 0.50\\
59 & 0.70\\
50 & 1.00\\
40 & 1.50\\
33 & 2.00\\
25 & 3.01\\
21 & 4.01\\
17 & 5.00\\
\noalign{\smallskip}
\hline
\end{array}
$$
\end{table}

   \begin{figure*}
   \centerline{\includegraphics[width=18cm]{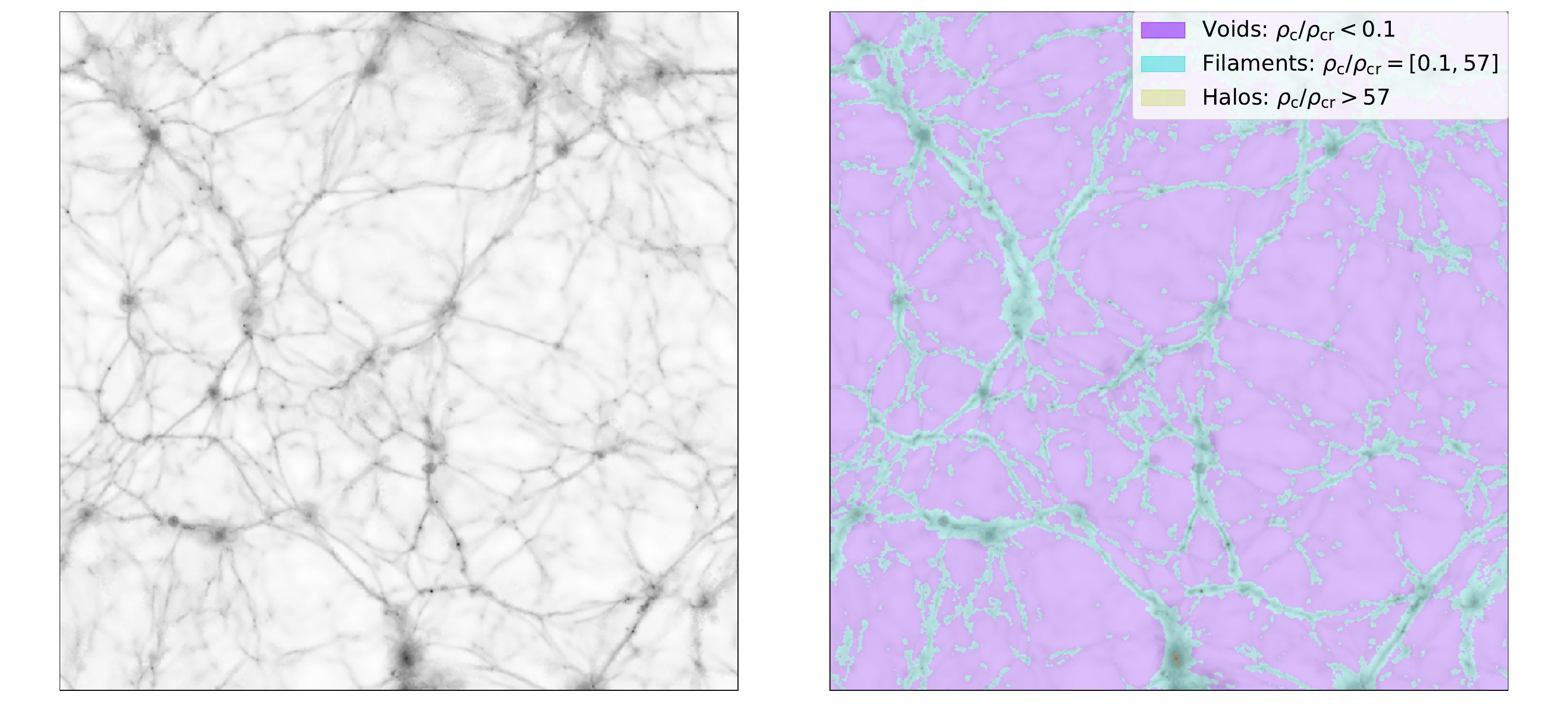}}
      \caption{Large-scale structure of the Cosmic Web. \textit{Left panel:} The density of gas (greyscale) lying along a two-dimensional $(75\,h^{-1}{\rm cMpc})^2$, infinitely thin slice across the $z=0$ snapshot of {\tt TNG100-3}. \textit{Right panel:} The same data, separated into halo (yellow), filament (cyan) and void (purple) substructures according to the metric used in this work.}
         \label{fig:Struct_Examples}
   \end{figure*}

In general, a cosmological hydrodynamic simulation can be treated as a three-dimensional box of evolving universe, which is initialised at an early time with a specific set of initial conditions and certain cosmological parameters. This box of universe is then evolved forward in time until reaching the present day $t_{0}$. At specific redshift intervals, instances of the simulation commonly referred to as `snapshots' are captured and stored for study. Individual snapshots are comprised of data (e.g., average mass density, star formation rate, dark matter density) recorded for smaller sub-volumes (hereafter `cells') of the simulation box. These cells dictate simulation resolution and can be fixed, or variable in size (e.g. in grid or particle -based simulations respectively), or a hybrid of the two (e.g. in so-called `moving-mesh' simulations). The recorded physical parameters for these cells are hereafter referred to as `fields'. For a review of simulation outputs, please refer to~\citet{humm17a}.

For FRB studies, various simulations such as the {\tt Magneticum Pathfinder Simulation}~\citep{dola15a} and the {\tt MICE Onion Simulation}~\citep{pol19a} have been used to estimate cosmological DM contributions. Simulations can also help to analyse DM contributions from host and foreground galaxies (e.g. {\tt RAMSES}, \citealt{zhu21a}). Other uses include FRB progenitor identification~\citep{dola15a}, cosmological parameter estimation (e.g. {\tt Illustris}, \citealt{jaro19b}) and the inference of the strength and origins of magnetic fields (e.g. {\tt CRPROPA},~\citealt{hack19a,hack20a}). In this work, we analyse FRB DM contributions from the Cosmic Web using {\tt IllustrisTNG}, a state-of-the-art cosmological hydrodynamic simulation suite.

\subsection{IllustrisTNG overview}
{\tt IllustrisTNG}~\citep{nels19a,pill18a,spri18a,Naim18a,nels18a,mari18a} is a moving-mesh cosmological hydrodynamic simulation suite, composed of Voronoi cells of density-dependent resolution. The simulations include sub-grid models to account for star formation, radiative metal cooling, stellar and supermassive black hole feedback, and chemical enrichment from SNII, SNIa, and AGB stars. The sub-grid models are calibrated to reproduce observational constraints such as the galaxy stellar mass function at $z=0$, the cosmic star formation rate density, and the galaxy stellar size distribution, among others \citep[see][for further details]{pill18a}.

The suite assumes Planck 2015 cosmology and is evolved between $127<z<0$. In total, 100 snapshots are recorded for a given {\tt TNG} run. Of these, 20 snapshots between $12<z<0$ are fully recorded and provide complete field information for every cell in the simulation. The rest are recorded as `mini' snapshots and omit information about certain fields\footnote{The full list of snapshots, and available fields, can be found at \url{https://www.tng-project.org/data/docs/specifications/}}.

{\tt TNG} is composed of multiple runs, each differing in volume and/or resolution\footnote{A full list of available TNG runs may be found at: \url{https://www.tng-project.org/data/docs/background/} (see \citet{nels19a} for further details).}. Together, these runs span cosmological \citep{suns23a} to sub-galactic \citep{boec23a} spacial scales, making the suite suitable for investigating DM contributions due to various cosmic structures. The largest volume run, {\tt TNG300}, consists of a box of approximately $\sim(300\,\rm{cMpc})^3$ and contains the largest number galaxies~\citep{nels19a}, making it the most suitable for our purposes. In this work, we therefore utilise `full' snapshots recorded for {\tt TNG300-1}, the highest-resolution version of this run. The respective redshifts of these snapshots are provided in Table \ref{table:snaps}.

\subsection{The electron density in {\tt IllustrisTNG}}\label{sec:ne}

In Eq.\ (\ref{DMIGMeq}), the physical electron density $n_{\rm e}$ can be written as
\begin{eqnarray}
n_{\rm e} = \frac{3\Omega_{\rm b} H_0^2}{8\pi G m_{\rm p}}(1+z)^3f_{\rm d}(z)f_{\rm e}(z),
\label{ne_theory}
\end{eqnarray}
where $f_{\rm d}$ and $f_{\rm e}$ are the fraction of baryons in the IGM and fraction of free electrons respectively, and $m_{\rm p}$ is the proton mass. According to~\citet{walt18a}, $f_{\rm e}$ can be written as (see also, e.g.~\citealt{deng14a,zhan18a,walt19a,macq20a,zhan20b,batt21a}):
\begin{eqnarray}
f_{\rm e}(z) = \left[(1-Y_{\rm p})\chi_{\rm e,H}(z) + Y_{\rm p}\chi_{\rm e,He}(z)/2\right],
\label{eq:figm}
\end{eqnarray}
where $Y_{\rm p}\simeq 0.24$ is the primordial helium abundance, $(1-Y_{\rm p})\simeq 0.76$ is therefore the hydrogen abundance, and $\chi_{\rm e,H}$ and $\chi_{\rm e,He}$ are the redshift-evolving ionisation fractions of hydrogen and helium. 

In {\tt TNG300-1}, the electron density is not a directly provided field, but can be calculated \citep{jaro20a,zhan20a,zhan20b,taka21a}. Star-forming cells in {\tt TNG300-1} comprise a mixture of gas in both cold and warm phases~\citep{katz96a,spri03a,pakm18a}. Since only ionised matter contributes to DM, we only incorporate warm-phase gas into our calculations for these cells, with the assumption that this gas is completely ionised. Thus, for any given {\tt TNG300-1} cell, we calculate $n_{\rm e}$ as follows~\citep{zhan20b}:
\begin{eqnarray}
n_{\rm e} = (w\eta_{\rm e})X_{\rm H}\frac{\rho}{m_{\rm p}}(1+z)^{3},
\label{ne_TNG}
\end{eqnarray}
where $\eta_{\rm e}$ is the cell's fractional electron number density with respect to its total hydrogen number density, provided by the {\tt TNG300-1} ``ElectronAbundance'' field, $X_{\rm H}=(1-Y_{\rm p})$, and $\rho$ is the gas cell comoving mass density. The warm-phase gas mass fraction, $w$, is defined as
\begin{eqnarray}
w = 
\begin{cases}
1 - x, & \text{for star forming cells} \\
1, & \text{otherwise}
\end{cases},
\label{Eq1}
\end{eqnarray}
where the cold-phase gas mass fraction, $x$, is given by
\begin{eqnarray}
         x =  \frac{u_{\rm h} - u}{u_{\rm h}-u_{\rm c}}.
      \label{Eq1a}
\end{eqnarray}
Here, $u$ is the cell's thermal energy per unit of gas mass and is provided by the {\tt TNG300-1} ``InternalEnergy'' field. Likewise,  $u_{\rm h}$ and $u_{\rm c}$ are the thermal energy per unit mass of hot- (${T_{\rm h}}\sim10^7\,\rm{K}$) and cold- (${T_{\rm c}}\sim~10^3\,\rm{K}$) phase gas respectively \citep{mari17a,spri03a}. We can calculate these values from their respective temperatures using\footnote{\url{https://www.tng-project.org/data/docs/faq/}}:
\begin{eqnarray}
u(T) = \frac{k_{\rm B}T}{\mu(\gamma-1)}\times\frac{\rm{Unit\,Mass}}{\rm{Unit\,Energy}},
\end{eqnarray}
where $k_{\rm B}$ is the Boltzmann constant in CGS unit, $\gamma=5/3$ is the adiabatic index, ${\rm Unit\,Mass}/{\rm Unit\,Energy}=10^{10}$ in {\tt TNG300-1} units and $\mu$ is the mean atomic weight:
\begin{eqnarray}
\mu = \frac{4m_{\rm p}}{1 + (3 + 4\eta_{\rm e})X_{\rm H}}.
\end{eqnarray}


\section{Large-Scale Structure}\label{sect:3}
   
The matter distribution and evolution of the Universe depends on the interplay between cosmological expansion, gravitational collapse, and cooling and feedback processes \citep{springel2006}. The resulting LSS, known as the Cosmic Web, has been mapped using both galaxy redshift and peculiar velocity surveys (see e.g.~\citealt{gott05a, tull14a} respectively). The Cosmic Web can be broken down into sub-structures such as dense halos, connecting filaments, and under-dense voids \citep{marti19a}. Fig.\ \ref{fig:Struct_Examples} shows a slice of {\tt TNG100-3} at $z=0$ along with these sub-structures.

Cosmological hydrodynamic simulations predict that at low redshifts, most baryons reside in a `warm-hot intergalactic medium' (WHIM) phase. This phase is induced through gravitational shock-heating to $10^5$-$10^7$\,K temperatures and $\sim 50\rho_{\rm cr}\Omega_{\rm baryon}$  densities where $\rho_{\rm{cr}}=3H^{2}_{0}/8\pi G$ is the critical density of the Universe at $z=0$, and $G$ is the gravitational constant \citep{marti19a}. The mass fraction in this phase increases from $z\sim3$, may equal that of cold gas by $z\sim1$ \citep{cen99a,medl21a}, and make up the majority of missing baryons by $z=0$ \citep{cen06a,marti19a}. At $z=0$, most of the WHIM is predicted to be tied up in filaments \citep{breg07a,marti19a}. The exact evolution of these structures is sensitive to AGN feedback, the modelling of which is of active interest \citep{haid16a,taka21a}. Observational evidence for gas in filamentary structures between galaxies, within superclusters, and on larger (ten to hundred-Mpc) scales, is growing statistically via Sunyaev-Zel'dovich studies and X-ray emission (see e.g.,~\citealt{tani19a,tani19b,tani20a,tani20b}). 

   \begin{figure}
   \centering
   \includegraphics[width=9cm]{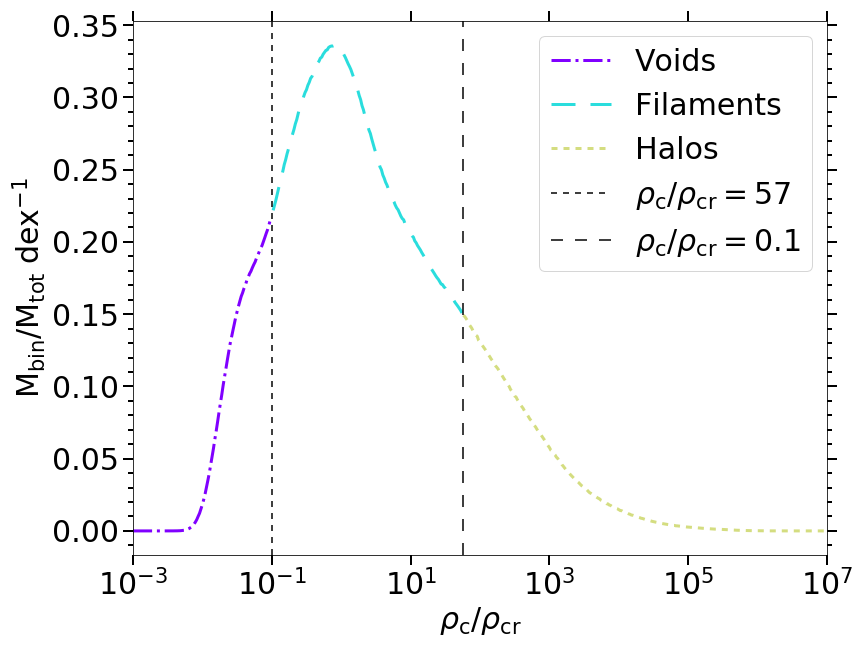}
      \caption{Mass distribution of large-scale structures. Following  \citet{arta21a}, the mass distribution within LSS substructures in the $z=0$ snapshot of {\tt TNG100-3} when binned according to our metric. Yellow, cyan, and purple portions of the curve indicate mass within halos, filaments, and voids respectively according to our boundaries (dashed and dot-dashed lines). $\rho_{\rm c}$ is the cold dark matter density, and $\rho_{\rm cr}$ is the present day's critical density of the Universe.}
         \label{fig:LSS_boundary_example}
   \end{figure}

Meanwhile, FRB observations have been used to confirm the existence of the missing baryons~\citep{macq20a}, and experiments have been proposed to constrain the amount of this matter tied up galaxies' CGM vs the diffuse IGM \citep{walt19a,ravi19a,lee21a}. Simulations have been used to study the DM contribution of the WHIM \citep{akah16a,medl21a}. It is therefore attractive to examine whether the DM contributions of LSS could prove to be a complementary tracer.

\subsection{LSS Classification in IllustrisTNG}\label{subsect:LSSclass}

   \begin{figure*}
   \centering
   \includegraphics[width=18cm]{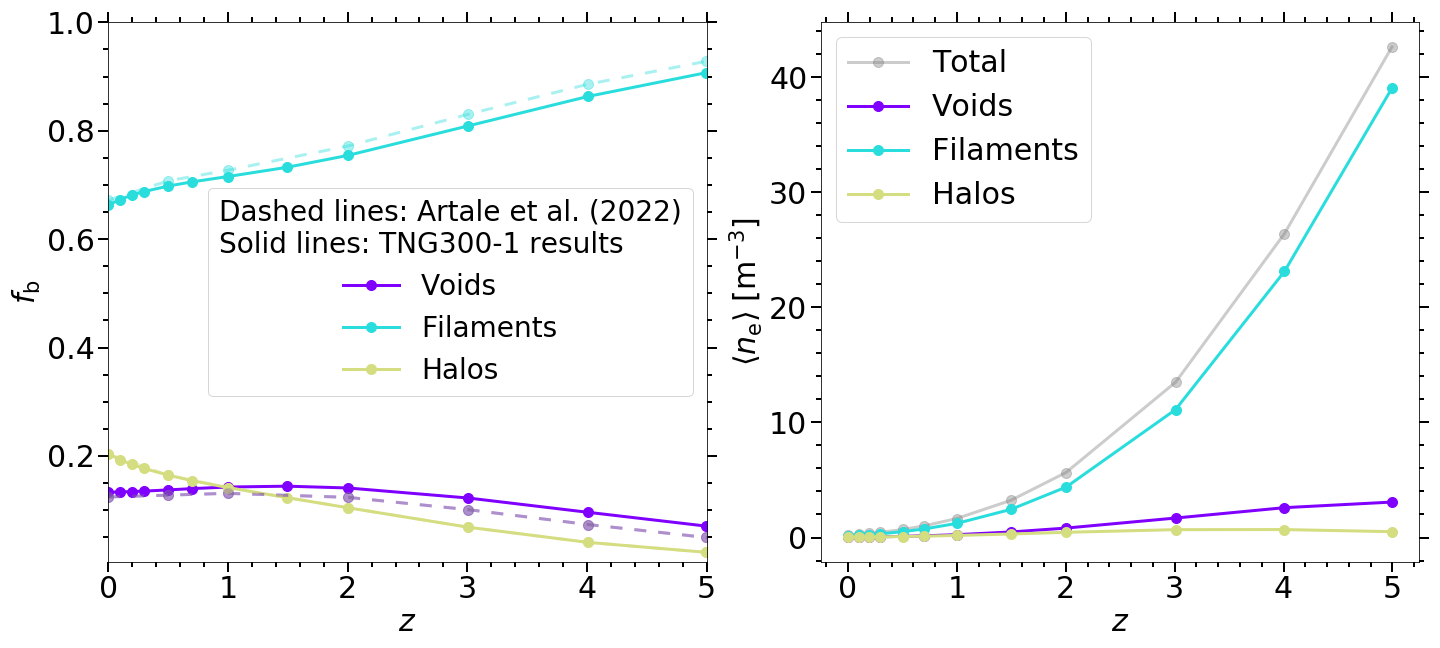}
      \caption{Matter evolution in large-scale structures. \textit{Left panel}: Evolution of the baryonic mass fraction $f_{\rm b}$ in halos (yellow), filaments (cyan) and voids (purple) as a function of redshift in {\tt TNG300-1} (solid lines), compared to the results of \citet{arta21a} (dashed lines). For halos, the solid and dashed lines overlap with each other. \textit{Right panel}: Evolution of the total mean electron density $\langle n_{\rm e}\rangle$ in {\tt TNG300-1} (grey line), and in each type of LSS (coloured lines), as a function of redshift.}
         \label{fig:LSSvsZ}
   \end{figure*}

Several different ways to classify LSS  are implemented in the literature (see \citealt{libe18a} for a review). Due to the influence of dark matter on LSS formation \citep{haid16a}, one simple method for identifying cosmic structures uses the local dark matter density of the simulation. We follow this method, as described by \citet{arta21a}, building upon on work by \citet{haid16a} and \citet{marti19a}, to classify {\tt TNG300-1} cells. According to this classification scheme, any given cell may be categorised as belonging to a halo, filament or void using the following metric:

      \begin{equation}
         \frac{\rho_{\rm{c}}}{\rho_{\rm{cr}}}{=}
         \begin{cases}
               < 0.1, & \text{for voids} \\
               0.1 - 57, & \text{for filaments}\\
               >57, & \text{for halos}\\
         \end{cases}
      \label{EqMetric}
      \end{equation}
where $\rho_{\rm{c}}$ is the cold dark matter density of the cell, provided by the {\tt TNG300-1} ``SubfindDMDensity'' field. The above halo-filament boundary is based on predictions from the Navarro-Frenk-White (NFW) models, and the filament-void boundary is selected to encompass the definition of halos and sheets according to each cell's gravitational potential \citep{arta21a,marti19a,fore09a}. A slice from the simulation coloured according to LSS type can be seen in Fig.\ \ref{fig:Struct_Examples}. Following \citet{arta21a}, Fig.\ \ref{fig:LSS_boundary_example} illustrates the amount of matter allocated to each type of structure in the simulation at $z=0$ according to our metric. Fig.\ \ref{fig:LSSvsZ} shows the evolution of baryonic mass and electron density within {\tt TNG300-1} as a function of LSS and redshift along with previous results by \citet{arta21a}.

\subsection{Further classification}\label{subsect:furtherclassification}

The practice of reconstructing matter densities along FRB sightlines is a burgeoning field of research. \citet{li19a}, \citet{proc19a} and \citet{simh20a} investigated the environments along sightlines to individual sources using information about spatially-proximate galaxies along with various matter density reconstruction techniques. \citet{lee21a} propose using similar methods for multiple sources to constrain CGM and IGM parameters. Therefore tracking the details of collapsed structures along sightlines in {\tt IllustrisTNG} is also prudent.

{\tt TNG} provides finely resolved details that are useful for this purpose in the form of ``sub-halos'', defined as virialised substructures within a dark matter halo \citep{wech18a}. In {\tt TNG}, sub-halos are classified using the {\sc subfind} algorithm \citep{spri01a,nels19a}. All {\tt TNG}-simulated galaxies are associated with a sub-halo. Through sub-halo analysis, any galaxies in proximity to FRB sightlines may therefore be identified. While sub-halo identification is not directly provided for individual {\tt TNG} cells via any field, the ``sub-halo ID'' for a given cell may be reconstructed by combining the cell's ``ParticleID'' field, the entire {\tt TNG} sub-halo catalog, and information supplied by relevant {\tt TNG} sub-halo offset tables\footnote{{\tt TNG} offset file information may be found at: \url{https://www.tng-project.org/data/docs/specifications/##sec3a}}. 

Not all {\tt TNG} cells are associated with sub-halos. These cells are flagged with a Sub-halo\;ID$=-1$. Additionally, sub-halos do not always contain galaxies. Non-galaxy sub-halos include sub-galactic, low-mass clumps of baryonic matter, which may be formed via, e.g., disk instabilities, which can be distinguished via a flag \citep{nels19a}. They may also be differentiated by enforcing cell requirements, e.g. containing non-zero stellar mass, or a certain number of stellar particles\footnote{\url{https://www.tng-project.org/data/forum/topic/235/how-to-identify-subhalos-containing-well-formed-ga/}}. These requirements may vary between {\tt TNG} runs. In this work, where appropriate, we classify collapsed structures both by considering all sub-halos, and by separating these sub-halos into bins according to their masses. We note, therefore, that statistics derived in upcoming sections which consider sub-halos of any mass may be influenced by environments containing some sub-halos which are small and not well resolved.

\section{Methods}\label{sect:4}

As with electron density, the ability to trace sightlines through {\tt IllustrisTNG} is not directly provided. However, multiple techniques to approximate the environments along sightlines exist. In principle, simulation-agnostic absorption line tools (e.g. {\tt TRIDENT}, \citealt{humm17a}) should be adaptable for use with {\tt IllustrisTNG}. In practice, we adapt here the methods first described in \citet{zhan20b}, whereby FRB sightlines are assembled from segments created from each simulation snapshot. To create a given segment, a random sub-volume is extracted from the total simulation volume, a line of sight defined, and the cells closest to this sightline identified. The ``true'' line of sight of the FRB is then approximated to propagate through these cells. Section \ref{subsect:pipes} summarises this process, and the process of obtaining average electron density and structural information for our segments. Calculating the DM accumulated along an individual segment is described in Sect.\ \ref{subsect:segmentanalysis}. Finally, Sect.\ \ref{subsect:LoS} describes the process of combining segments to study different DM contributions out to cosmological distances. 

\subsection{Line of sight segments}\label{subsect:pipes}

\begin{table}
\caption[]{The {\tt IllustrisTNG} fields drawn upon in this work. Uses include electron density and DM calculation, LSS classification, sub-halo identification and impact parameter analysis. The unit ``${\rm ckpc}$'' refers to ``comoving kiloparsec'', a comoving quantity \citep{nels19a}.}
\label{table:fields}
$$
\begin{array}{cc}
\hline
\noalign{\smallskip}
\text{Field} & \text{TNG\,units}\\
\noalign{\smallskip}
\hline
\text{Coordinates} & h^{-1}{\rm ckpc}\\
\text{Density} & \frac{10^{10}h^{-1}{\rm M}_{\odot}}{(h^{-1}{\rm ckpc})^3}\\
\text{ElectronAbundance} & -\\
\text{InternalEnergy} & {({\rm km}\,{\rm s}^{-1})^2}\\
\text{Masses} & 10^{10}h^{-1}{\rm M}_{\odot}\\
\text{ParticleIDs} & -\\
\text{StarFormationRate} & {\rm M}_{\odot}\;{\rm yr}^{-1}\\
\text{SubfindDMdensity} & \frac{10^{10}h^{-1}{\rm M}_{\odot}}{(h^{-1}{\rm ckpc})^3}\\
\noalign{\smallskip}
\hline
\end{array}
$$
\end{table}

Following \citet{zhan20b}, we generate a single line of sight segment for a given {\tt TNG300-1} snapshot as follows. Initially, header data detailing snapshot size and redshift is loaded, alongside a matchlist which allows simulation cells to be related to their parent sub-halos, as discussed in Sect.\ \ref{subsect:furtherclassification}. The segment's physical dimensions are then defined. A random Cartesian coordinate of the form ($0,y,z$) is generated as starting point of the line of sight through the segment. The segment's rectangular extent, of ({205,0.2,0.2})\,$h^{-1}{\rm cMpc}$ where the $x$-dimension spans the entire length of the simulation box, is selected so that the segment extends out to 0.1\,$h^{-1}{\rm cMpc}$ either side of the origin in the $y$ and $z$ directions. The corresponding end-point of the sightline will be ($205,y,z$)\,$h^{-1}{\rm cMpc}$.

Once the segment coordinates are defined, all relevant simulation fields are loaded or otherwise created for this snapshot. For clarity, these fields are listed in Table \ref{table:fields}. Warm and cold-phase gas mass fractions are calculated as described in Sect.\ \ref{sec:ne}. Field values along the segment sightline are obtained from the cells which lie closest to 10,000 linearly spaced, $\sim20\,h^{-1}{\rm ckpc}$-wide bins defined to span the length of the sightline. The LSS type associated with each bin is defined as discussed in Sect.\ \ref{subsect:LSSclass}. The number of cells along the segment sightline belonging to each structure type are calculated. Electron densities of every cell along the segment sightline are calculated as discussed in Sect.\ \ref{sec:ne}. These are averaged to obtain the average total electron density traversed along the segment. The average electron densities associated with each LSS type across the segment sightline are also calculated separately.

\begin{figure*}
   \centering
   \includegraphics[width=16cm]{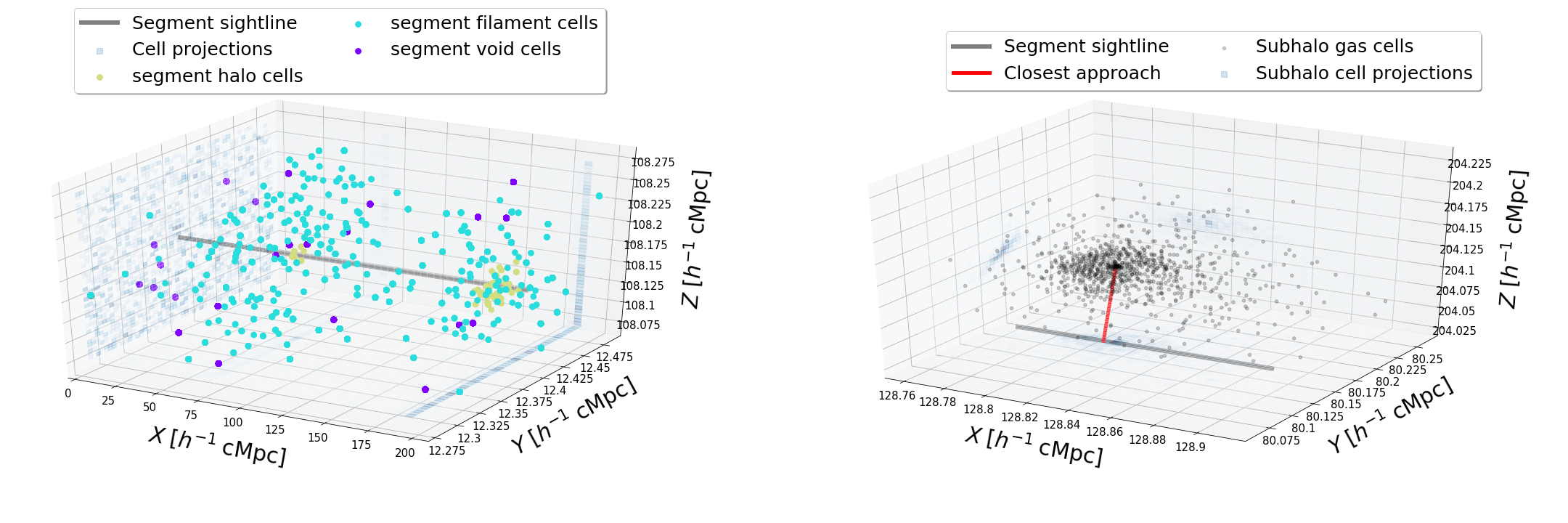}
      \caption{Tracing structure along sightlines. \textit{Left panel}: A ({205,0.2,0.2})\,$h^{-1}{\rm cMpc}$ line of sight segment traversing a {\tt TNG} snapshot at $z=0$. The grey line indicates the sightline through the segment. Blue squares projected onto the ($xy,yz,xz$) planes represent the number density of all cells within the segment in the ($x,y,z$) directions. Dots, coloured according to their LSS type, represent the central locations of all cells identified as closest to one of the 10,000 bins along the sightline. The sightline is approximated to traverse these cells due to the Voronoi tessellation of the simulation. \textit{Right panel}: A second sightline, complete with a sub-halo it traverses and the impact parameter measured between the sightline and sub-halo.}
         \label{fig:pipeLSS}
   \end{figure*}

Additionally, Sub-halo IDs associated with cells along the sightline are identified as discussed in Sect.\ \ref{subsect:furtherclassification}. Unique sub-halos are stored, and all bins along the segment associated with these sub-halos are identified. In this way, all DM attributed to these sub-halos may also be calculated. The sightline's coordinates of closest approach to the central positions of these sub-halos are also stored for further analysis. An example segment can be seen in the left panel of Fig.\ \ref{fig:pipeLSS}. An example of impact parameter analysis can be seen in the right panel of Fig.\ \ref{fig:pipeLSS}.

\subsection{DM within a segment\label{subsect:segmentanalysis}}

Using the information recorded for each segment, the DM accumulated along its line of sight can be calculated. We compute the total DM along a given segment sightline by converting the continuous Eq.\ (\ref{eq:DM_def}) into a discrete sum over its bins assuming a negligible redshift change over the length of a snapshot:

    \begin{equation}
    \begin{aligned}
        {\rm DM}=\widetilde{\rm DM}/(1+z),
    \end{aligned}
    \label{eq:discreteDMnoncosmic}
    \end{equation}
where $\widetilde{\rm DM}$ is the total DM along the segment's sightline in its rest frame,

\begin{eqnarray}
\widetilde{\rm DM}=\sum^{10,000}_{n=1}\frac{{\rm d}\,{\rm DM}}{{\rm d}l}\Big|_{n}\times {\rm d}l.
\label{eq:discreteDMnoncosmic2}
\end{eqnarray}

In Eq.\ (\ref{eq:discreteDMnoncosmic2}), $({\rm d\,DM}/{\rm d}l)_{n}=n_{\rm e}$ is the physical electron density of the $n^{\rm th}$ bin as described by Eq.\ (\ref{ne_TNG}), and ${\rm d}l=a{\rm d}\eta$ is the physical distance increment of propagation through the bin, where $a$ is the scale factor at the redshift of the segment, and $d\eta$ is the comoving distance increment, equal to the bin width in comoving {\tt IllustrisTNG} units\footnote{For more information on converting {\tt TNG} coordinates, see: \url{https://www.tng-project.org/data/docs/specifications/}}. It thus follows that the DM accumulated by a limited portion of a sightline traversing a particular sub-halo can be calculated by constraining Eq.\ (\ref{eq:discreteDMnoncosmic2}) to only those bins associated with its specific Sub-halo ID.

\subsection{Combining DM from different segments}\label{subsect:LoS}

Following \citet{zhan20b}, to form complete sightlines between sources at redshift $z$ and an observer at $z=0$, we exploit the lack of redshift evolution within snapshots, approximating them as infinitesimally narrow redshift slices in order to convert Eq.\ (\ref{DMIGMeq}), the continuous theoretical integral for $\rm{DM_{cos}}$, into a discrete cumulative sum over our snapshots which we denote $\rm{DM^{TNG}_{total}}$:

{\begin{eqnarray}
    \label{eq:discrete}
{\rm DM^{TNG}_{total}}(z_{i+1}) &=& {\rm DM^{TNG}_{total}}(z_i) \nonumber \\
   &+& \frac{1}{2}\left(\frac{{\rm d}\,{\rm DM_{TNG}}}{{\rm d}z}\biggr\rvert_{z_i}+\frac{{\rm d}\,{\rm DM_{TNG}}}{{\rm d}z}\biggr\rvert_{z_{i+1}}\right)(z_{i+1}-z_{i}), \nonumber \\
\end{eqnarray}}
where $z_i$ is the redshift of each snapshot, ${\rm DM^{TNG}_{total}}(z)|_{z=0}=0$, and for a given segment of a given snapshot,

    {\begin{equation}\label{eq:integrand}
        \frac{{\rm d}{\rm DM_{TNG}}}{{\rm d}z}\biggr\rvert_{z=z_i} = \frac{cn_{\rm e}(z_i)}{H_0(1+z_i)^2\sqrt{\Omega_{\rm m}(1+z)^3+\Omega_{\Lambda}}},
    \end{equation}}
where $n_{\rm e}$ is the average electron density computed as described in Sect.\ \ref{subsect:pipes}. Additionally, if any cell along the segment sightline is associated with a dark matter sub-halo (see Sect.\ \ref{subsect:furtherclassification}) then we consider the sub-halo to have been traversed by the sightline. By recording the number of unique sub-halos which are traversed by each segment, the number of sub-halos $N_{\rm sub}$ traversed as a function of redshift for a single sightline may then be calculated similarly to Eq.\ (\ref{eq:discrete}) using: 

\begin{eqnarray}
N_{\rm sub}(z_{i+1}) &=& N_{\rm sub}(z_i) \nonumber \\
        &+& \frac{1}{2}\left(\frac{{\rm d}N_{\rm sub}}{{\rm d}z}\biggr\rvert_{z_i}+\frac{{\rm d}N_{\rm sub}}{{\rm d}z}\biggr\rvert_{z_{i+1}}\right)(z_{i+1}-z_{i}),
    \label{eq:discrete_subhalo}
\end{eqnarray}
with the initial condition $N_{\rm sub}(z)|_{z=0}=0$, and where ${\rm d}N_{\rm sub}/{\rm d}z$ is the number of sub-halos traversed by a given segment.

Although full {\tt TNG300-1} snapshots extend out to $z=12$, and FRBs may be detectable by current instruments out to at least $z=10$ \citep{zhan18a}, the ionisation information which informs our electron densities may be inaccurate at $z>6$ \citep{nels18a}. We thus generate segments as discussed in Sect.\ \ref{subsect:pipes} for only the snapshots listed in Table \ref{table:snaps}, which extend out to $z=5$. We calculate $N_{\rm sub}$ and the cosmological $\rm{DM^{TNG}_{total}}$ for these, and calculate the portions of DM acquired only from individual LSS types ($\rm{DM^{TNG}_{halos}}$, $\rm{DM^{TNG}_{filaments}}$, $\rm{DM^{TNG}_{voids}}$) by substituting $n_{\rm e}$ in Eq.\ (\ref{eq:integrand}) with the average electron densities associated only with their corresponding structures. Following \citet{zhan20b}, we create a total of 5125 unique segments of dimensions ({205,0.2,0.2})\,$h^{-1}{\rm cMpc}$ for each snapshot; these are randomly sampled to create 10,000,000 individual FRB lines of sight. We also observe the convention of constructing our FRB sightlines along the $x$-axes of our snapshots due to computing constraints \citep{jaro19b,zhan20b}. Including sightlines constructed along all three axes would better sample the entire simulation, and hence improve results. In principle, adapting the absorption line tool {\tt TRIDENT} \citep{humm17a} for use with {\tt IllustrisTNG}, perhaps in conjunction with techniques discussed in \citet{batt21a} for bridging gaps between simulation snapshots, could prove another powerful way to study not just average, but individual sightlines across the full breadth of the {\tt IllustrisTNG} simulation suite.

   \begin{figure*}
   \centering
   \includegraphics[width=18cm]{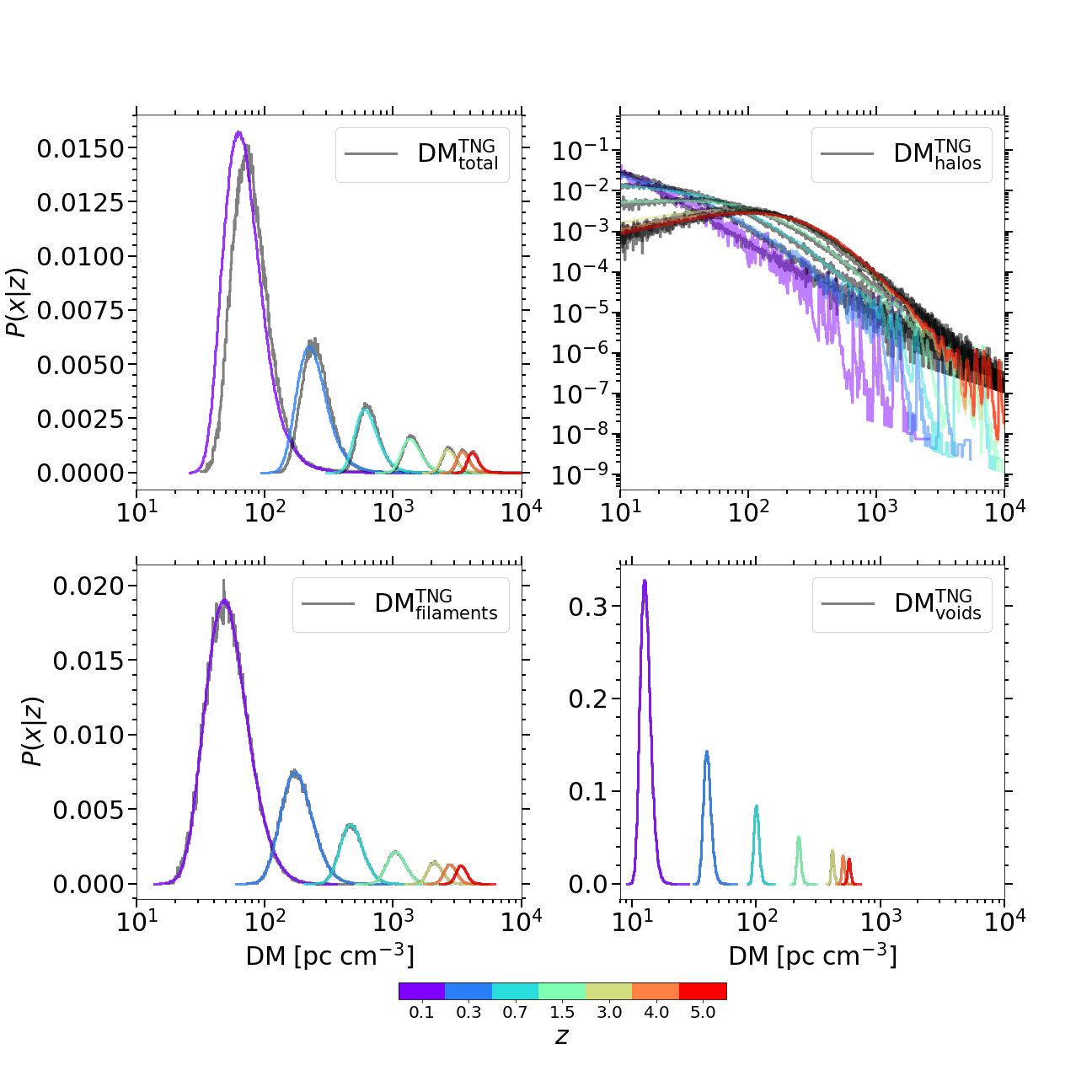}
      \caption{FRB DM evolution with redshift. \textit{Clockwise from top left}: Histogrammed probability distributions (solid coloured lines) of the total cosmological DM ($\rm{DM^{TNG}_{total}}$), and contributions by large-scale structures ($\rm{DM^{TNG}_{halos}}$, $\rm{DM^{TNG}_{filaments}}$, $\rm{DM^{TNG}_{voids}}$) for FRBs originating at various redshifts (see Table \ref{table:snaps} and colourbar), calculated for {\tt TNG300-1}. For comparison, grey lines are new data drawn from fits to these LSS distributions. These new data are combined to create the comparison histograms for $\rm{DM^{TNG}_{total}}$. The slight mismatch between our low-redshift TNG data and drawn ${\rm DM^{TNG}_{total}}$ values arises due to our fits slightly overestimating the probability of sightlines containing very high ($\gtrsim10^3\,{\rm pc\;cm^{-3}}$) ${\rm DM_{halos}^{TNG}}$ contributions at low redshifts.}
         \label{fig:LSSDMbreakdown}
   \end{figure*}

   \begin{figure*}
   \centering
   \includegraphics[width=18cm]{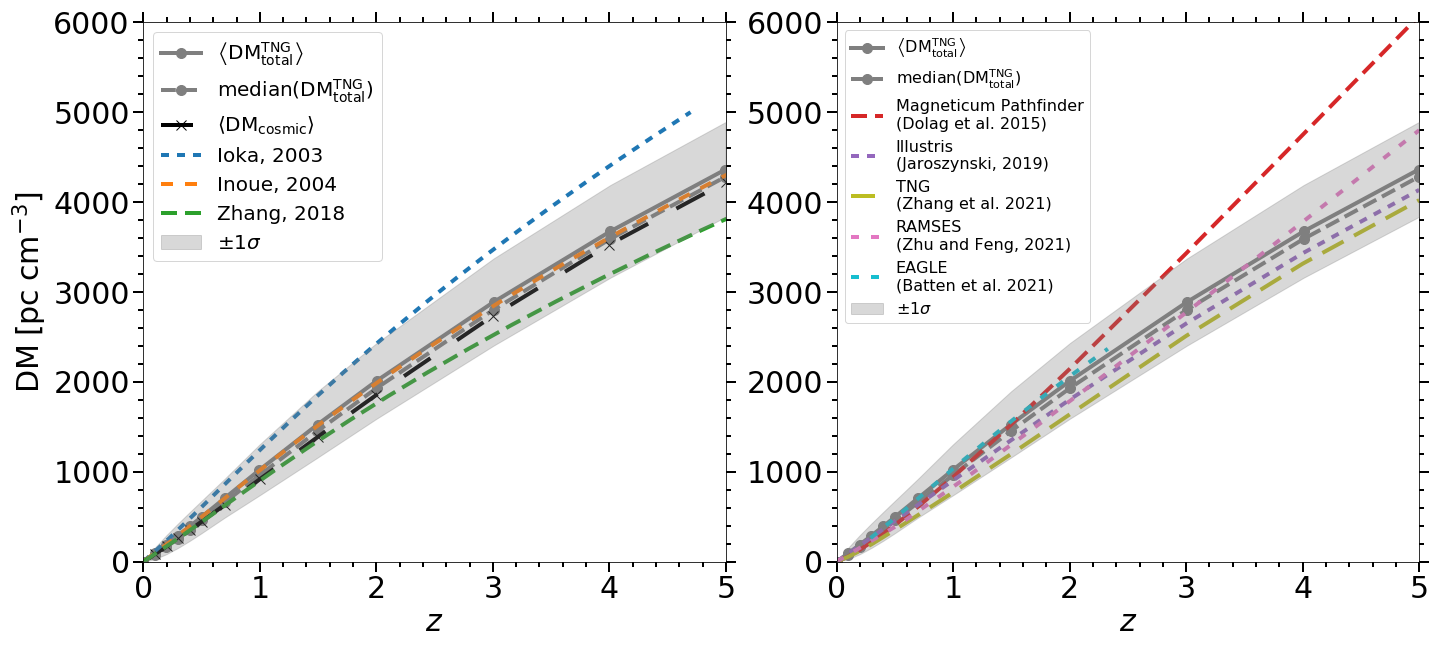}
      \caption{FRB DM evolution with redshift continued. \textit{Both panels:} The average total DM-$z$ relationship derived for {\tt TNG300-1}. The mean (solid grey line), median (dotted grey line) and standard deviation (shaded grey region) around the mean of $\rm{DM^{TNG}_{total}}$ values calculated for our sightlines are shown. \textit{Left panel:} These are compared to $\langle\rm{DM_{cosmic}}\rangle$ (dashed black line), and the results of analytical studies from previous literature (dashed coloured lines). \textit{Right panel:} Compared to results derived from other cosmological simulations in previous literature (dashed coloured lines).}
         \label{fig:LSSstats}
   \end{figure*}

Figure \ref{fig:LSSDMbreakdown} presents normalised histograms showing the evolution of the distribution of $\rm{DM^{TNG}_{total}}$ values calculated for our sightlines, and for their LSS contributions, as a function of redshift. We find that $\rm{DM^{TNG}_{filaments}}$ and $\rm{DM^{TNG}_{voids}}$ are both best fit by lognormal distributions\footnote{\url{https://docs.scipy.org/doc/scipy/reference/generated/scipy.stats.lognorm.html}} of the form

\begin{eqnarray}
P(y,a)=\frac{1}{s}\frac{1}{ay\sqrt{2\pi}}\exp\left(-\frac{\ln^2(y)}{2a^2}\right),
    \label{eq:fil_void_fit_function}
\end{eqnarray}
and that $\rm{DM^{TNG}_{halos}}$ can be approximated using the positive range of a log-logistic distribution\footnote{\url{https://docs.scipy.org/doc/scipy/reference/generated/scipy.stats.fisk.html}} fitted to the natural logarithm of the DM data:

\begin{eqnarray}
P(y,a)=\frac{1}{s}\frac{ay^{a-1}}{(1+y^a)^2}.
  \label{eq:halo_fit_function}
\end{eqnarray}

In each case, $a$ refers to the distribution's  shape parameter, and

    \begin{equation}
    \begin{aligned}
        {y}=\frac{(x-l)}{s} \phantom{-1 <{}} & \text{where}
                 \begin{cases}
               {x} = \text{DM}, & \text{for voids, filaments} \\
               {x} = \ln({\text{DM}}), & \text{for halos}\\
         \end{cases}
    \end{aligned}
    \label{eq:hal_fil_void_fit_function2}
    \end{equation}
    with shifting parameters $l$ and scaling parameters $s$ respectively. The best-fit parameters for each distribution at each redshift are provided in Table \ref{table:results2}. Histograms of data drawn from these fits are included in Fig.\ \ref{fig:LSSDMbreakdown} for comparison purposes. The draws from the void and filamentary fits are a good match to the TNG data distributions, while draws from the the halo fits are relatively good match, but slightly overestimate the probability of low-redshift sightlines containing high ($\gtrsim10^3\,{\rm pc\;cm^{-3}}$) ${\rm DM^{TNG}_{halo}}$ contributions.

Using our distributions, we calculate various statistics (mean, median, standard deviation) for the relationship between $\rm{DM^{TNG}_{total}}$ and redshift. In Fig.\ \ref{fig:LSSstats}, we compare these to similar results reported in previous literature. These comparisons are further discussed in Sect.\ \ref{sect:othersims}. In Fig.\ \ref{fig:LSSfracts} (left), we compare the average behaviours of $\rm{DM^{TNG}_{halos}}$, $\rm{DM^{TNG}_{filaments}}$ and $\rm{DM^{TNG}_{voids}}$. We present more detailed breakdowns of the fractional contributions to FRB DMs by each type of large-scale structure as a function of redshift in Fig.\ \ref{fig:LSSfracts} (right), and provide computed values at our redshift intervals for all measurements in Table \ref{table:results}.

\section{Results}\label{sect:5}

In previous sections, we have reviewed the ingredients and methods required to approximate both LSS and DM for IllustrisTNG. Together, these elements may be combined to provide detailed insight into the redshift-evolving impact of cosmological large-scale structures on FRB signals. In this Section, we present the results of our analysis.

\subsection{LSS analysis}

Figure \ref{fig:LSSDMbreakdown} presents individually normalised distributions for $\rm{DM^{TNG}_{total}}$ values which are accumulated along our sightlines by FRBs originating at various redshifts. Individually normalised distributions for ($\rm{DM^{TNG}_{halos}}$, $\rm{DM^{TNG}_{filaments}}$, $\rm{DM^{TNG}_{voids}}$), the contributions to the total DM by each type of LSS, are also shown. The similarity between the total, and filamentary DM distributions (Fig.\ \ref{fig:LSSDMbreakdown} top left, bottom left) is immediately apparent, indicating that, for most sightlines, the filamentary contribution dominates DM. The filament and halo distributions also display the widest DM ranges, and high-DM tails, including for FRBs originating at low redshifts. This behaviour indicates that the ionised electron distributions in halo and filament environments may significantly vary between FRB sightlines, and that in rare instances, an FRB may propagate through a region of much higher electron density, which will contribute large amounts of observed DM. In contrast, the narrower distributions of the void components indicates that voids contribute a smaller range of possible DM values to FRB sightlines, and thus that voids may remain generally more homogeneous environments from sightline to sightline.

\begin{figure*}
   \centering
   \includegraphics[width=18cm]{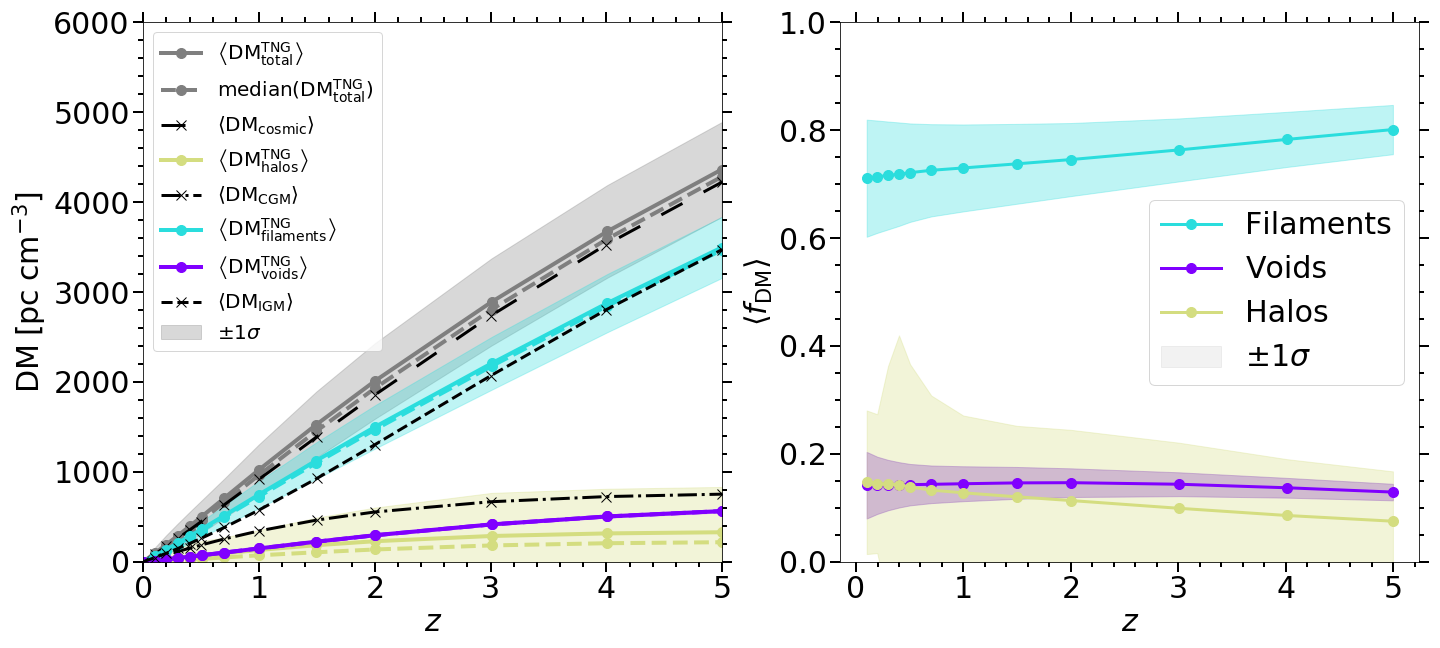}
      \caption{Redshift evolution of DM contributions as a function of large-scale structure. \textit{Left panel:} The mean, median, and standard deviation around the mean of the DM-redshift relationship for {\tt TNG300-1} halos ($\rm{DM^{TNG}_{halos}}$; yellow), filaments ($\rm{DM^{TNG}_{filaments}}$; cyan), and voids ($\rm{DM^{TNG}_{voids}}$; purple), derived using our metric. These are compared to the $\langle\rm{DM_{cosmic}}\rangle$ subcomponents (see Sect.\ \ref{sect:othersims}): $\langle\rm{DM_{IGM}}\rangle$ (dashed black line) and $\langle\rm{DM_{CGM}}\rangle$ (dot-dashed black line). \textit{Right panel}: The means (coloured lines) and standard deviations (shaded regions) of the equivalent fractional DM contributions $\langle f_{\rm DM}\rangle$ accumulated due to traversing each type of structure by our FRBs originating at given redshifts.}
         \label{fig:LSSfracts}
   \end{figure*}

The differing behaviours of the contributions to DM by each type of LSS as a function of redshift are quantified using various statistics presented in Fig.\ \ref{fig:LSSfracts} and Table \ref{table:results}. In Fig.\ \ref{fig:LSSfracts} (left), we compare the evolution of the average total observed DM from TNG, $\langle{\rm DM^{TNG}_{total}}\rangle$, to the evolution of its LSS subcomponents, $\langle{\rm DM^{TNG}_{halos}}\rangle$, $\langle{\rm DM^{TNG}_{filaments}}\rangle$, and $\langle{\rm DM^{TNG}_{voids}}\rangle$. Here, the dominance of the filamentary contribution to the average FRB sightline is more apparent. It can be seen that filaments contribute an increasingly large portion of DM to the total for FRBs originating at higher redshifts, while halos and voids contribute to a lesser extent. We find that for FRBs originating at $z=0.1$ vs $z=5$, the average contributions to DM from halos, filaments, and voids will grow from $\sim[13.64/65.82/13.13]\,{\rm pc\;cm^{-3}}$ to $\sim[327.61/3494.53/563.02]\,{\rm pc\;cm^{-3}}$ respectively. The standard deviations of these components quantify the DM variability due to each type of LSS along our sightlines. It can be seen that the largest DM scatter arises due to material in halos, followed by filaments, and that voids have very little scatter. A more detailed breakdown of the evolution of these average components can be found in Table \ref{table:results}. 

In Fig.\ \ref{fig:LSSfracts} (right), we quantify the relative evolution of our LSS DM components as a function of redshift using their average fractional contributions to the total DM. We again see that $\langle\rm{DM^{TNG}_{filaments}}\rangle$ is the dominant contribution to $\langle\rm{DM^{TNG}_{total}}\rangle$, and that, on average, the fraction of DM accumulated from filaments, $\langle f^{\rm DM}_{\rm filaments}\rangle$, rises from $\sim71\%$ to $\sim80\%$ for an FRB originating at $z=0.1$ vs $z=5$. Meanwhile, the average fraction accumulated from halos, $\langle f^{\rm DM}_{\rm halos}\rangle$, halves, from $\sim15\%$ to $\sim8\%$. The average fractional contribution from voids, $\langle f^{\rm DM}_{\rm voids}\rangle$, remains relatively consistent, to within $\sim1\%$, regardless of source redshift. Here, too, the significant scatter in $\rm{DM^{TNG}_{halos}}$ and $\rm{DM^{TNG}_{filaments}}$ can be seen, and all calculated values can be found in Table \ref{table:results}. These results support previous predictions that denser structures, such as galaxy groups and clusters, may be major culprits of any observed sightline-to-sightline variance in FRB DMs \citep{proc19a,zhu21a,conn22a}.

\subsection{Sub-halo analysis}

\begin{figure*}
   \centering
   \includegraphics[width=18cm]{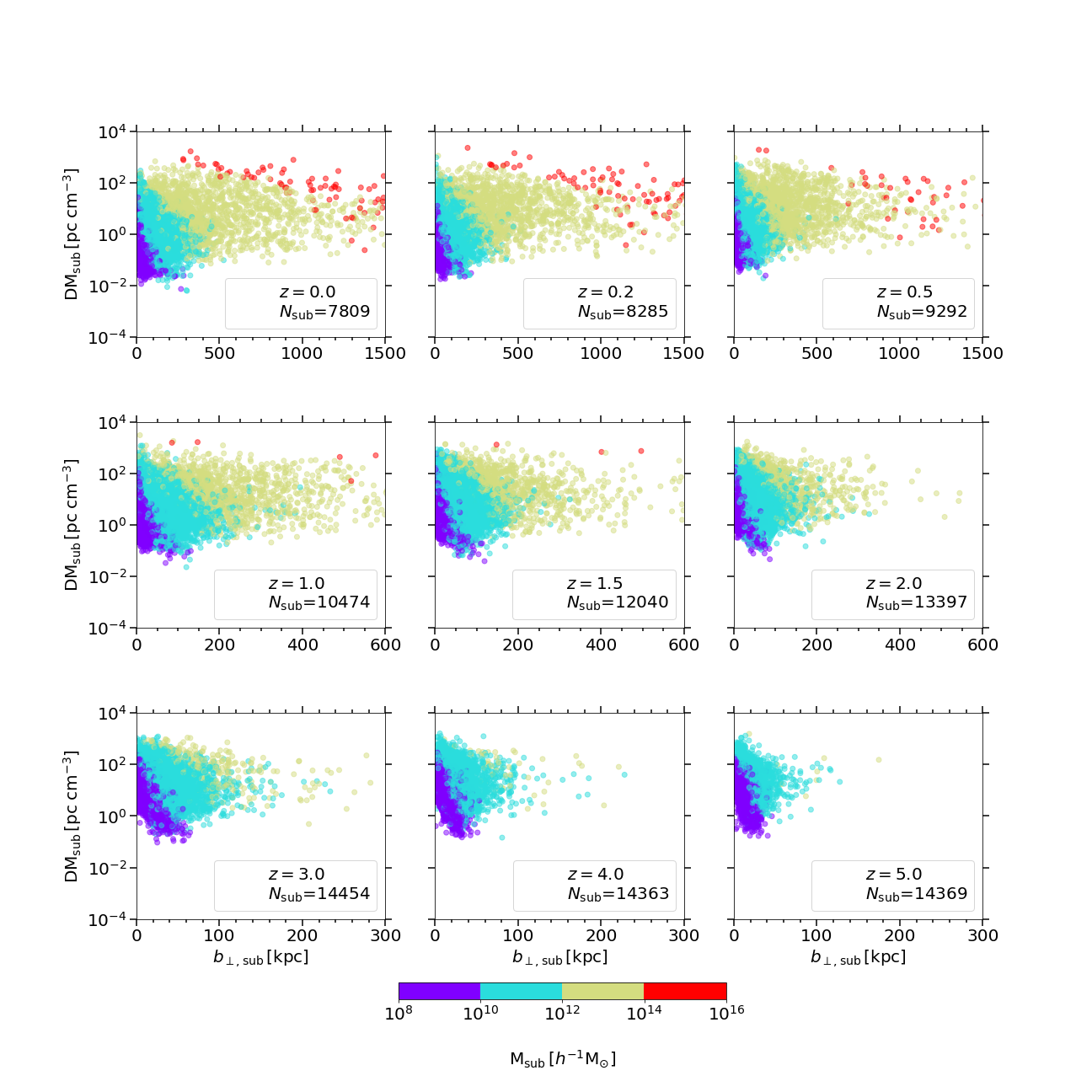}
      \caption{Collapsed structures traversed by our segments. \textit{Each panel:} Scatter points indicating the distributions in impact parameter ($b_{\perp,{\rm sub}}$) - accumulated dispersion measure (${\rm DM_{sub}}$) parameter space for any sub-halos considered traversed by our segments, in the snapshot of redshift $z$ described in the legend. Each scatter point colour is determined using its sub-halo mass $M_{\rm sub}$, according to the mass bin ranges described by the colour bar. The total number of sub-halos $N_{\rm sub}$ traversed by all segments \textit{within a given snapshot} is provided in the legend.}
         \label{fig:DMvIF}
   \end{figure*}

Table \ref{table:segmentsubhalos} quantifies the number of unique sub-halos which are intersected by our segments at various redshifts. We provide the total number of sub-halos intersected for a given snapshot, and bin these according to their masses. Fig.\ \ref{fig:DMvIF} visualises these data, along with the impact parameters $b_{\perp,\,\rm{sub}}$ and accumulated dispersion measures ${\rm DM_{sub}}$ which are associated with these intersections. It is immediately apparent that the general trend is towards fewer sub-halo intersections at lower redshifts, presumably due to the merging of lower mass sub-halos to form higher mass sub-halos over cosmic time. We can also see that larger potential impact parameters are possible for low-redshift sub-halos, presumably due to increased sizes. In addition, for a given redshift, higher mass sub-halos generally contribute larger values of ${\rm DM_{sub}}$, and allow for larger potential impact parameters, indicating potentially larger radii. We anticipate that the exact cutoffs in ${\rm DM_{sub}}$-$b_{\perp,\,{\rm sub}}$ -space for sub-halos of different mass bins results from the radius-density profile of these sub-halos.

\begin{figure*}
   \centering
   \includegraphics[width=18cm]{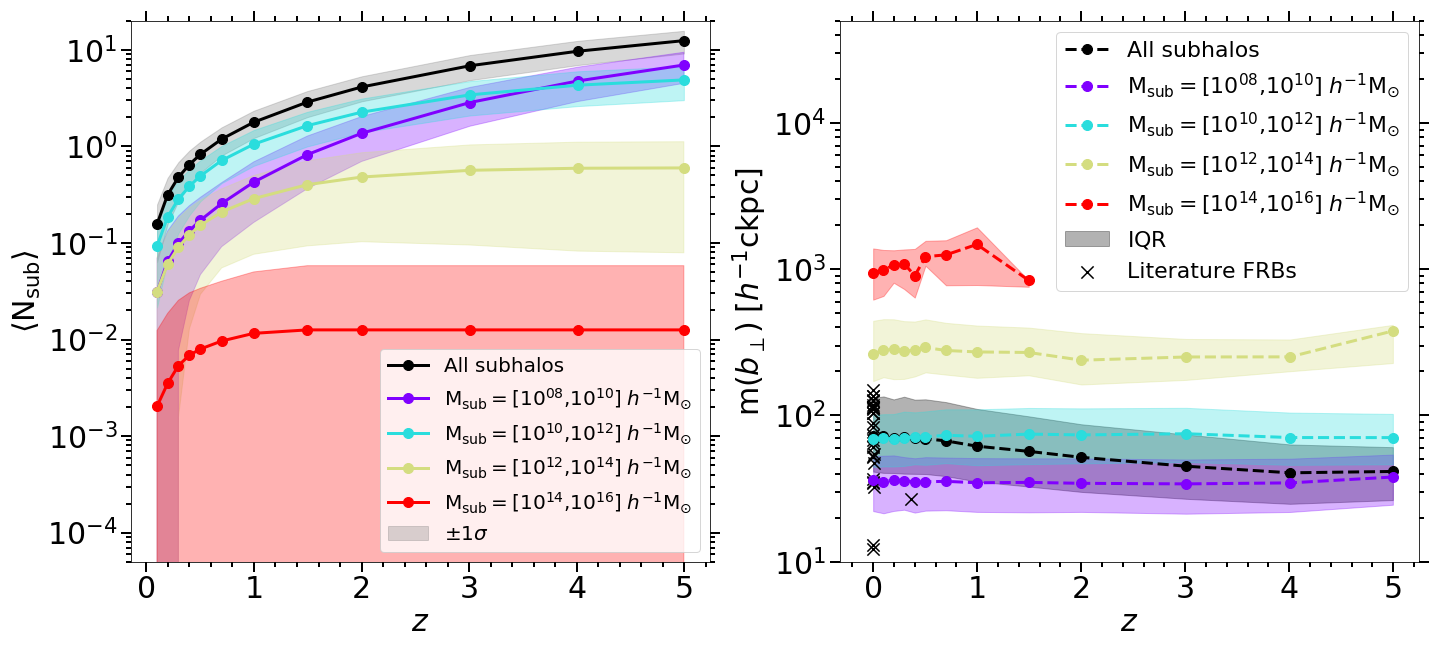} 
      \caption{Collapsed structures traversed with redshift. \textit{Left panel}: Statistics (mean, standard deviation) describing the average number of sub-halos which are traversed by FRB sightlines out to given redshifts within {\tt TNG300-1}. Black lines and grey shaded regions describe the total number of sub-halos traversed; coloured lines and shaded regions describe the number of sub-halos when binned by mass according to the mass bins of Fig.\ \ref{fig:DMvIF}. \textit{Right panel}: Statistics (median, interquartile range) describing the average impact parameters between our sightlines and these structures within each snapshot. Overplotted crosses are the impact parameters between observed FRBs and galaxies as calculated by \citet{conn22a}, \citet{conn20a}, and \citet{proc19b}.}
         \label{fig:substats}
   \end{figure*}

We quantify the average evolution of these intersections for complete FRB sightlines as a function of redshift in Fig.\ \ref{fig:substats} and Tables \ref{table:resultsTraversed} and \ref{table:resultsImpact}. Fig.\ \ref{fig:substats} (left) and Table \ref{table:resultsTraversed} detail the average number of sub-halos that an FRB originating at a given redshift will traverse according to Eq.\ (\ref{eq:discrete_subhalo}). We present the mean and standard deviations calculated for our 10,000,000 sightlines. Fig.\ \ref{fig:substats} (right) and Table \ref{table:resultsImpact} detail the average impact parameters which we measure between our segments and these sub-halos at the redshifts of our snapshots. Due to the non-Gaussianity which we observe for these distributions, we present median, interquartile range (IQR) and interpercentile range (IPR) values calculated for these data. For comparison, we overplot derived positions in impact parameter - redshift space for a number of observed FRBs which are considered likely to have intersected galaxies including M31 and M33, according to \citet{proc19b}, \citet{conn20a} and \citet{conn22a}. Where not otherwise provided by the literature, we have converted the galaxies' impact parameters from physical to comoving distances using their preferred redshifts, which we obtained using the NASA Extragalactic Database (NED). Where a galaxy was reported to have a blueshift, $z=0$ was used in the conversion process. In both Fig.\ \ref{fig:substats} subplots, we present the statistics obtained when considering all sub-halos traversed by our sightlines, and those we obtain when considering sub-halos binned by mass according to the bins in Fig.\ \ref{fig:DMvIF}. We note that we only provide impact parameter statistics for a given mass bin and redshift in Table \ref{table:resultsImpact} when the total number of sub-halos traversed by segments in the snapshot for this mass bin is $\ge40$ (see Table \ref{table:segmentsubhalos}). 

The left panel of Fig.\ \ref{fig:substats} shows that, on average, FRBs originating beyond $0.5<z<0.7$ will intersect at least one sub-halo of any mass during propagation. The total number of sub-halos which are on average traversed will increase from $\left<N_{\rm sub}\right>\simeq1.8$ for an FRB originating at $z=1$, to $\left<N_{\rm sub}\right>\simeq12.4$ for an FRB originating at $z=5$. By studying these intersected sub-halos according to their masses, one sees that the traversed sub-halos are dominated by those of mass $M_{\rm sub}=[10^8,10^{12}]\,h^{-1}{\rm M_{\odot}}$. On average, less than one sub-halo of mass $M_{\rm sub}\geq10^{12}\,h^{-1}{\rm M_{\odot}}$ will be traversed, even by an FRB originating at $z=5$. When comparing the relative behaviours of the mass-binned halos as a function of redshift, one sees that FRBs originating at $z\lesssim3$ are more likely to traverse higher-mass sub-halos ($M_{\rm sub}=[10^{10},10^{12}]\,h^{-1}{\rm M_{\odot}}$), but for FRBs originating at $z\gtrsim4$, the most commonly intersected collapsed structures are of lower mass ($M_{\rm sub}=[10^8,10^{10}]\,h^{-1}{\rm M_{\odot}}$), with the transition occurring at around $z\sim3.8$. This transition in the encounter rates of lower vs higher mass sub-halos likely stems from the population of less massive sub-halos dwindling at later times due to continuous mergers.

By studying the impact parameters between our sightlines and mass-binned traversed sub-halos in Fig.\ \ref{fig:substats} (right), it can be seen that the average impact parameter between our sightlines and sub-halos increases with mass, from a median $b_{\perp,{\rm sub}}\simeq36.1\,h^{-1}{\rm ckpc}$ for sub-halos of mass $M_{\rm sub}=[10^8,10^{10}]\,h^{-1}{\rm M_{\odot}}$, to $b_{\perp,{\rm sub}}\simeq950\,h^{-1}{\rm ckpc}$ for sub-halos of mass $M_{\rm sub}=[10^{14},10^{16}]\,h^{-1}{\rm M_{\odot}}$. These average impact parameters remain relatively consistent for sub-halos with masses $M_{\rm sub}=[10^8,10^{14}]\,h^{-1}{\rm M_{\odot}}$, regardless of the redshift of intersection. However, the average impact parameter measured when considering sub-halos of any mass falls after $z>1$. This is presumably due to a deficit of sub-halos of the highest masses ($M_{\rm sub}>10^{14\,}h^{-1}{\rm M_{\odot}}$) at higher redshifts, which have not yet had time to form within the {\tt TNG300-1} box.

Finally, we quantify the average DMs which are accumulated by during the traversal of the sub-halos in Fig.\ \ref{fig:subDMs} and Table \ref{table:resultsDM}. Fig.\ \ref{fig:subDMs} (left) details the average dispersion measure which will be accumulated by traversing any sub-halos at a given redshift in the rest frame of the sub-halos, $\widetilde{\rm DM}_{\rm sub}$ (see also, Sect.\ \ref{subsect:segmentanalysis}). Fig.\ \ref{fig:subDMs} (right) and and Table \ref{table:resultsDM} detail the average DM which will be \textit{observed} due to traversing sub-halos at these redshifts. Here, too, we present statistics obtained when considering traversed sub-halos of any mass, and when considering sub-halos binned by their masses. It can be seen that although $\widetilde{{\rm DM}}_{\rm sub}$ increases with redshift in all cases, this increase is greatly suppressed when considering observed DM, due to the $(1+z)^{-1}$ factor. Again, we only provide DM statistics for a given mass bin and redshift in Table \ref{table:resultsDM} when the total number of sub-halos traversed by segments in the snapshot for this mass bin is $\ge40$ (see Table \ref{table:segmentsubhalos}).

\section{Discussion}\label{sect:6}

The study of matter distributions in difficult-to-observe phases is a burgeoning field of research. Observationally, Sunyaev-Zel'dovich and X-ray studies are already being used \citep{tani19a,tani19b,tani20a,tani20b}; as are cosmological hydrodynamic simulations, to within the constraints afforded by, e.g., their feedback models. In the future, large galaxy catalogs provided by surveys such as LSST \citep{ivez19a} might also be useful in conjunction with matter density reconstruction techniques (see, e.g., \citealt{wang09a,wang12a,wang16a}) for reconstructing large volumes of LSS. FRBs, with their isotropic sky distribution \citep{petr22a} and the insight they provide into the ionised environments which they traverse, may yet prove a valuable way to augment such techniques, and aid our understanding of the ionised matter distribution of the Universe.

 In view of this, we have investigated the evolution with redshift of the cosmological DM contribution; the contributions of its constituent halos, filaments and voids; the likelihood of intercepting foreground collapsed structures; and the impact parameters and DM contributions associated with these interceptions, for FRBs within {\tt TNG300-1}. In this Section, we compare our results to previous investigations utilising different analytical, empirical and theoretical methods, different simulations and LSS classification schemes, and real FRB observations.

\subsection{LSS analysis}\label{sect:othersims}

In Fig.\ \ref{fig:LSSstats}, we compare $\langle\rm{DM^{TNG}_{total}}\rangle$, the average total DM that the cosmological component will contribute to an FRB originating at a given redshift according to our simulation, to similar results from previous literature. In most cases, the results appear reasonably consistent, with many remaining within one standard deviation of the {\tt TNG} result out to  $z=4$.

Using the publicly available \texttt{FRUITBAT} repository \citep{batt21a}\footnote{\url{https://github.com/abatten/fruitbat}} we reproduce the results of \citet{ioka03a}, \citet{inou04a} and \citet{zhan18a} in Fig.\ \ref{fig:LSSstats} (left). These are analytical estimates which approximate Eq.\ (\ref{ne_theory}), and thus DM, using different IGM baryon fractions, ${f_{\rm d}}$, and free electron fractions, ${f_{\rm e}}$, and assume these to be constant with redshift (see \citealt{batt21a} for a brief review). These values evolve with redshift in {\tt TNG} (see Eq.\ (\ref{ne_TNG})), which will account for some of the differences between the analytical results and our analysis. Conversely, the evolution of $\langle\rm{DM_{cosmic}}\rangle$ with redshift (see, e.g., \citealt{proc19a, macq20a}), which we recreate using the public {\tt FRB} github repository\footnote{\url{https://github.com/FRBs/FRB}} for Fig.\ \ref{fig:LSSstats} (left), is derived by combining theoretical predictions and empirical techniques, and allows for redshift-evolving IGM baryon and ionisation factors. It appears much closer to our result. 

We also compare our results to the most probable DM-$z$ relationships obtained using various cosmological simulations in Fig.\ \ref{fig:LSSstats} (right). These include the medium-resolution {\tt Magneticum Pathfinder simulation} \citep{dola15a}, the original {\tt Illustris} simulation \citep{jaro19b}, the {\tt RAMSES} simulation \citep{zhu21a}, the {\tt EAGLE} simulation \citep{batt21a}, and {\tt IllustrisTNG}, when interpolating the results of \citet{zhan20b}. It is notable that our recreation of the \citet{zhan20b} result (see Fig.\ \ref{fig:LSSstats} (right), yellow dashed line), which we achieve by interpolating their fits, yields systematically lower DMs as a function of redshift than our own result, despite both being derived using the same run of the same simulation ({\tt TNG300-1}). We posit that these differences arise for two reasons. Firstly, a systematic shift towards lower DMs in the peak of the \citet{zhan20b} fitted distributions compared with their data distributions can be seen when comparing each for a given redshift. Secondly, \citet{zhan20b} elect to discard cells containing measurable star formation during their DM calculations, whereas we apply a correction factor (see Eq.\ (\ref{Eq1})) to these cells, and still include some portion of their material into our calculations. Minor differences between our result and the majority of the other simulations likely arise due to technical differences and underlying assumptions which inform the simulations. These could include the sizes, resolutions, and AGN feedback models of the simulations, and the exact values of the cosmological parameters which they are initialised with.

All considered, the majority of the other curves lying within one standard deviation of our result demonstrates a broad consistency with previous literature, albeit highlighting the non-negligible effect that underlying cosmological parameters and galaxy evolution mechanisms have on the true DM-redshift relationship.

In the left panel of Fig.\ \ref{fig:LSSfracts}, we compare the evolution of $\langle\rm{DM^{TNG}_{total}}\rangle$ to the evolution its large-scale structure subcomponents, $\langle\rm{DM^{TNG}_{halos}}\rangle$, $\langle\rm{DM^{TNG}_{filaments}}\rangle$, and $\langle\rm{DM^{TNG}_{voids}}\rangle$. For reference, we plot these against another interpretation of the cosmological DM component. In this Figure, $\langle\rm{DM_{cosmic}}\rangle$, and its subcomponents, $\langle\rm{DM_{CGM}}\rangle$\footnote{Referred to as $\langle\rm{DM_{halos}}\rangle$ in the github repository, but renamed here to avoid confusion with our definition of $\langle\rm{DM^{TNG}_{halos}}\rangle$} and $\langle\rm{DM_{IGM}}\rangle$, are calculated using the public {\tt FRB} github repository. The value $\langle\rm{DM_{CGM}}\rangle$ describes the average DM  contributed to $\langle\rm{DM_{cosmic}}\rangle$ by ionised material residing within the CGM of any galactic halos which lie along FRB sightlines as a function of redshift, and is computed using the Aemulus Halo Mass Function \citep{mccl19a}. Thus the value $\langle\rm{DM_{IGM}}\rangle$, which is defined as $\langle\rm{DM_{cosmic}}\rangle - \langle\rm{DM_{CGM}}\rangle$, must attribute any remaining DM to matter within the diffuse IGM, but outside of foreground galactic halos. When compared with our derived values, it can be seen that $\langle\rm{DM_{CGM}}\rangle$ consistently accounts for a larger portion of $\langle\rm{DM_{cosmic}}\rangle$ than $\langle\rm{DM^{TNG}_{halos}}\rangle$ contributes to $\langle\rm{DM^{TNG}_{total}}\rangle$. This likely results from the Aemulus Halo Mass Function incorporating all matter classified as belonging to halos according to our metric, as well as some of the denser matter lying closer to these halos, which we might classify as belonging to filaments. Likewise, it can be seen that $\langle\rm{DM_{IGM}}\rangle$ consistently contributes a smaller portion to $\langle\rm{DM_{cosmic}}\rangle$ than $\langle\rm{DM^{TNG}_{filaments}}\rangle$ contributes to $\langle\rm{DM^{TNG}_{total}}\rangle$, presumably because it includes only the less dense portion of the matter which we classify as filamentary, along with any matter we classify as belonging to voids.

\citet{zhu21a} have previously used the {\tt RAMSES} simulation to investigate the DM contributions of large-scale structures between $0<z<1$, assuming different definitions for large scale structure types. They report values at $z=0$, which we can compare to our results at $z=0.1$. Assuming equivalence between \citet{zhu21a}'s definition of nodes and our definition of halos, and that our definition of filaments encompasses \citet{zhu21a} filaments and walls, their results translate to $\sim23.5\%$, $\sim61.4\%$ and $\sim15\%$ of cosmological DM being attributed to halos, filaments, and voids respectively at $z=0$. These values are consistent to within one standard deviation of our results at $z=0.1$.

Conversely, \citet{akah16a} have previously studied the the contributions to DM by large-scale structures out to $z\sim5$ using the $\Lambda$CDM universe simulations. They adopt a different classification scheme, defining their structures to be voids, sheets, filaments (which they also define as the WHIM) and clusters according to gas temperature ($T<10^4\,{\rm K}$; $10^4\,{\rm K}< T < 10^5\,{\rm K}$; $10^5\,{\rm K}< T < 10^7\,{\rm K}$; and $T >10^7\,{\rm K}$ respectively). Their work yields different results when compared with our Table \ref{table:results}. While the fractional contributions to total DM by matter which we classify as halos, and \citet{akah16a} classify as clusters, both decrease with redshift, our filamentary fraction grows between $0.1<z<5$, and our void fraction decreases. In \citet{akah16a} these two contributions evolve in the opposite direction. \citet{akah16a} voids increase their relative DM contribution with redshift, overtaking the DM contributed by their filaments by $z\sim2$. By $z\sim5$, voids exceed filaments in their fractional contribution to total DM by a factor $>2$, and by over $\sim1000\,\rm{pc\,cm^{-3}}$, which is more than all matter which we classify as voids ever contributes. Assuming that our definition of filaments encompasses the \citet{akah16a} definitions of both filaments and sheets (see, e.g., \citealt{marti19a}, who show that filaments defined using local dark matter density are equivalent to filaments plus sheets when defined using the tidal tensor method) may reduce this discrepancy to some extent. However this still does not reconcile the two sets of results. As with our comparison of $\langle\rm{DM^{TNG}_{total}}\rangle$ to previous literature, the root of the discrepancy between the trends for our LSS contributions likely derives from fundamental differences between our simulations (e.g. underlying galaxy feedback models), along with our different methods for constructing FRB sightlines and our categorisation of LSS according to gravitational collapse rather than temperature.

When analysing the relative contributions to DM accumulated from our different LSS types, the relatively low variance in the DM contributed by voids between sightlines presents an interesting notion. \citet{walt18a} has previously shown that the constraining power of FRBs on cosmological parameters is limited by the inhomogeneity of the IGM from sightline to sightline, and that minimising this variance may be necessary in order to use FRBs to measure the dark energy equation of state. It thus follows that, could a sample of FRBs propagating along low-structure sightlines traversing only (or mostly) void material be assembled, these low-variance sources might offer the ability to obtain tighter cosmological constraints than by using the full FRB population. This idea has been touched on by \citealt{bapt23a}, who, building on the $P({\rm DM}|z)$ model of \citealt{macq20a}, parameterised the halo gas contribution to its deviation from a Gaussian in terms of a fluctuation parameter, $F$. While investigating the degeneracy of this parameter with $H_0$, they note that the low DM end of the $P({\rm DM}|z)$ distribution is less affected by the choices of $F$ and $H_0$, implying that lower-structure sightlines (e.g. through voids) may better constrain $H_0$.

As the measurable parameter for observed FRBs is total DM, to obtain such a sample, one would have to inventively select probable low-structure sightline FRBs from the overall population. Potential methods might include selecting well-localised FRBs with observed DMs which are lower than expected when considering their $P{\rm (DM|}z)$ redshift distributions (see, e.g., \citealt{walk20a}), or using, e.g., photometric techniques \citep{simh21a} to identify structureless sightlines. Such work could also potentially help constrain the baryon distribution within voids.

   \begin{figure*}
   \centering
   \includegraphics[width=18cm]{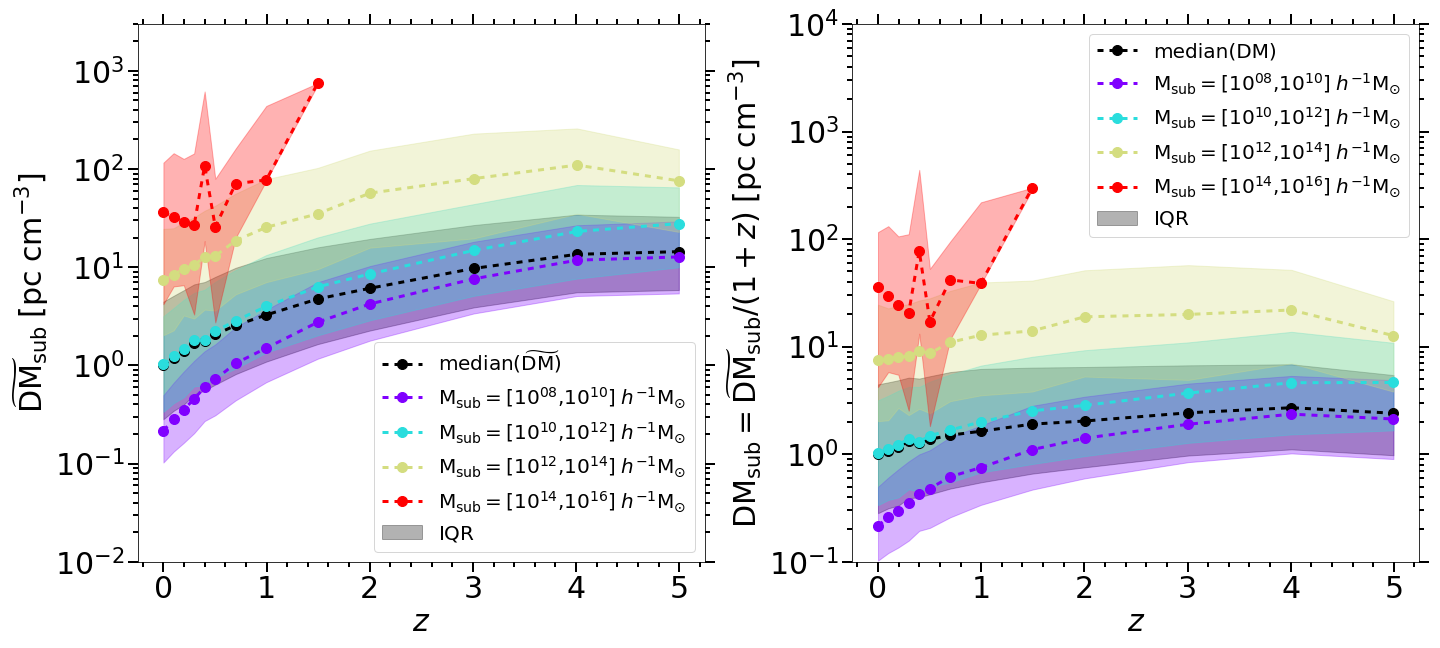}
      \caption{DM accumulated within individual snapshots due to traversing collapsed structures. \textit{Left panel:} Statistics (median, interquartile range) describing the average ${\rm \widetilde{DM}}$ accumulated within snapshots by sightlines traversing our sub-halos in the rest frames of the snapshots, and thus, the rest frames of the sub-halos. \textit{Right panel:} These values weighted by redshift, effectively describing the average DM which would be observed by an observer at $z=0$ due to these sub-halos. Colours indicate the mass ranges described in Fig.\ \ref{fig:substats}.}
         \label{fig:subDMs}
   \end{figure*}

\subsection{Sub-halo analysis}\label{sect:IFanalysis}

In Fig.\ \ref{fig:substats} (left) and Table \ref{table:resultsTraversed}, we quantify the average number of collapsed structures traversed by an FRB originating at a given redshift, according to our simulations. In Fig.\ \ref{fig:substats} (right) and Table \ref{table:resultsImpact}, we quantify the average impact parameters $b_{\perp}$ which we measure between any collapsed structures traversed by our sightlines, and the sightlines which traversed them, at the redshifts of our snapshots. For reference, we have compared these results to real observational FRB data. \citet{proc19b} have studied the halo gas of a galaxy intersected by an ASKAP-localised FRB, and \citet{conn20a} have reported on a single FRB which intersects regions of both M33 and M31. Additionally, \citet{conn22a} discuss unlocalised, non-repeating, CHIME-detected FRBs which they deem likely to have intersected galaxies (but unlikely to have passed directly through their disks) during their propagation through the IGM. The FRB-galaxy intersection discussed by \citet{proc19b} occurs at $z\sim0.37$ with the $\sim10^{12}\,h^{-1}{\rm M_{\odot}}$ galaxy FG-181112 at $b_{\perp}\sim27\,h^{-1}{\rm ckpc}$. This impact parameter lies outside of the IQR ($=[39.88,128.27]\,h^{-1}{\rm ckpc}$) which we measure at $z=0.4$ when considering impact parameters between all sub-halos and our sightlines, and when considering impact parameters to sub-halos of mass $M_{\rm sub}=[10^{12},10^{14}]\,h^{-1}{\rm M_{\odot}}$ ($=[184.85,438.88]\,h^{-1}{\rm ckpc}$). It additionally lies outside of the 10th to 90th percentiles ($=[112.76,663.54]\,h^{-1}{\rm ckpc}$) which we measure for intersections with sub-halos of mass $M_{\rm sub}=[10^{12},10^{14}]\,h^{-1}{\rm M_{\odot}}$, though it lies within these percentiles when considering sub-halos of any mass. These results support \citet{proc19b}'s analysis that this is a particularly rare intersection event (which they place at a probability of $\sim0.5\%$). 

Of the 28 FRB-galaxy impact parameters discussed by \citet{conn20a} and \citet{conn22a}, all are associated with galaxies which lie at $z\simeq0$. Of these 28 impact parameters, we find that 19 lie within the IQR ($=[41.33,131.86]\,h^{-1}{\rm ckpc}$) which we measure when considering the impact parameters between sub-halos of all masses and our sightlines at $z=0$. Additionally, we find that 26 of these 28 lie between the 10th and 90th percentiles ($=[23.06,303.08]\,h^{-1}{\rm ckpc}$) which we measure for our impact parameters when considering sub-halos of all masses. The FRB-galaxy impact parameters which lie outside these ranges are associated with FRB20190430B \citep{conn22a} and FRB191108 \citep{conn20a}, in their purported intersections with the galaxies NGC6015 and M33 at $b_{\perp}\simeq12.9\,h^{-1}{\rm ckpc}$ and $b_{\perp}\simeq12.2\,h^{-1}{\rm ckpc}$ respectively. As demonstrated by \citet{proc19b}, FRBs may impact intervening galaxies at closer distances on rare occasions, thus we see no discrepancy between our analysis and this result. However, we note that every galaxy considered by \citet{conn22a} is reported to with $M_{\rm vir}>10^{12}\,h^{-1}{\rm M_{\odot}}$, placing them in our $[10^{12},10^{14}]\,h^{-1}{\rm M_{\odot}}$ sub-halo mass bin. Many of the \citet{conn22a} $b_{\perp}$ values fall outside our measured ranges when considering sub-halos in this mass bin alone.

Figure \ref{fig:subDMs} and Table \ref{table:resultsDM} quantify the average dispersion measures accumulated within individual snapshots by our sightlines due to traversing these sub-halos. Fig.\ \ref{fig:subDMs} (left) presents the average dispersion measures $\widetilde{{\rm DM}}_{\rm sub}$ which are accumulated within snapshots of given redshifts by our segments due to traversing their sub-halos, in the rest frame of the sub-halos. We find these increase with redshift, both when considering all traversed sub-halos, and when considering sub-halos binned by mass. Fig.\ \ref{fig:subDMs} (right) presents the average \textit{observed} dispersion measures ${\rm DM_{sub}}$ due to these intersections. We find the observed DMs to be strongly suppressed as a function of redshift by the $(1+z)^{-1}$ factor. These trends generally align well with previous literature which examines the DM contributions of FRB host galaxies, e.g. \citet{jaro20a}, \citet{zhan20a}, and (Kov\'{a}cs et al., in prep.), who also report increasing, albeit suppressed, average ${\rm DM_{host}}$ contributions as a function of redshift. The works report systematically larger average $\rm{DM_{host}}$ contributions (of order tens to hundreds of ${\rm pc\;cm^{-3}}$ \citealt{jaro20a,zhan20a}) than our average ${\rm DM_{sub}}$ values (which have medians of order ${\rm DM_{sub}}\simeq1\,{\rm pc\;cm^{-3}}$ for sub-halos of any mass, increasing for more massive sub-halos). This, however, is to be expected, given that FRBs are embedded in their host galaxies, and thus more likely to be exposed to denser material, including from their local environments, than when traversing foreground sub-halos.

While the evolution of the DM contribution due to collapsed structures as a function of redshift may be consistent with previous simulations, we find that, for a given redshift, the DM distributions themselves are highly skewed towards lower DMs, and that on average, the DM amplitudes which we observe do not align well with observational results. \citet{conn22a} report an average observed DM excess of $\sim 90\,\rm{pc\,cm^{-3}}$ for their likely galaxy-intersecting sources, compared to those unlikely to intersect galaxies. Their purported galaxies of intersection are all nearby ($z\simeq0$), with masses $\sim10^{12}\,h^{-1}{\rm M_{\odot}}$. The $\sim90\,{\rm pc\;cm^{-3}}$ DM excess lies outside of our measured IQR ($=[2.03,24.52]\,{\rm pc\;cm^{-3}}$) for $M_{\rm sub}=[10^{12},10^{14}]\,h^{-1}{\rm M_{\odot}}$ sub-halos, and also just above the range we measure between the 10th and 90th percentiles for these data ($=[0.56,69.09]\,{\rm pc\;cm^{-3}}$). Indeed, the reported DM excess lies above the interquartile ranges, and 90th percentiles which we measure for ${\rm DM_{sub}}$ for all but the most massive ($M_{\rm sub}>10^{14}\,h^{-1}{\rm M_{\odot}}$) TNG sub-halos (see Table \ref{table:resultsDM}).

When considering results derived from {\tt TNG} sub-halos of any mass, some non-galaxy sub-halos could potentially skew statistics towards lower DMs (see Sect.\ \ref{subsect:furtherclassification}). Equally, however, some of this DM discrepancy could be attributed to matter lying outside of sub-halos as defined by {\tt TNG}. \citet{conn22a} themselves note that their observational excess DM is unexpectedly large, citing expected means of $<40\,\rm{pc\,cm^{-3}}$. They combine their unexpectedly large DM excess, with the observation that many of the purportedly intersected galaxies reside within galaxy groups, to conclude that FRBs have the potential to become effective probes of diffuse ionised matter within dark matter halos, between galaxy groups and clusters; and to allow for the testing of models describing dark matter halos and the feedback which informs these matter distributions. Our results therefore provide a further incentive to continue investigating the DM contributions of cosmological LSS. As more FRBs are detected, well-localised, and observed to intersect galaxies at a range of redshifts, the statistics of these intersections may prove valuable for testing both {\tt IllustrisTNG}, and feedback models.

\section{Concluding remarks}\label{sect:7}

In this work, we have studied the redshift-evolving contributions to DM by large-scale and collapsed structures along FRB sightlines using {\tt TNG300-1}.

We find that filaments dominate the cosmological DM contribution, increasingly so for FRBs originating at larger redshifts. The filamentary contribution rises from $\sim71\%$ to $\sim80\%$ for FRBs originating at $z=0.1$ and $z=5$, respectively, while the halo contribution falls from $\sim15\%$ to $\sim8\%$, and the void contribution remains consistent to within $\sim1\%$. We find that the majority of the scatter in DM from sightline to sightline originates from halo and filamentary matter. Conversely, the DM contribution from voids varies less from sightline to sightline, indicating voids may be homogeneous environments. As it has been shown that constraining cosmological parameters using the DM-redshift relationship is limited by the variance in dispersion measures from sightline to sightline \citep{walt18a}, we posit that leveraging void-only, and other relatively low-structure sightlines may prove an effective method for probing void baryon distributions, and more precisely constraining cosmological parameters using FRBs.

We find that, on average, an FRB sightline will intersect one foreground collapsed dark matter structure, or sub-halo, of any mass by between $0.5<z<0.7$. An FRB originating at $z=1$ will on average intersect $\langle N_{\rm sub}\rangle \simeq1.8$ sub-halos, along its propagation path. FRBs originating at $z=5$ will on average intersect $\langle N_{\rm sub}\rangle \simeq12.4$ sub-halos. The impact parameters between our simulated sightlines and {\tt TNG} sub-halos of any mass appear consistent with those measured for purported galaxy-intersecting FRBs from the literature. We find that of 28 purported FRB-galaxy impact parameters at $z\sim0$ \citep{conn20a,conn22a}, 19 lie within our measured IQR ($=[41.33,131.86]\,h^{-1}{\rm ckpc}$) for impact parameters at $z=0$, and 26 lie within the 10th and 90th percentiles ($=[23.06,303.08]\,h^{-1}{\rm ckpc}$) of our data. The remaining two $z\sim0$ sightlines, and an FRB-galaxy intersection at $z\sim0.37$ \citep{proc19b} may simply be rarer cases of FRB sightlines more closely impacting galaxies along their propagation paths. However, we find higher $b_{\perp}$ values than observed for the purported intersected galaxies when considering a subset of {\tt TNG} sub-halos closer to their masses.

We find that in the rest frame of our traversed sub-halos, the average accumulated ${\rm DM_{sub}}$ increases with the redshift of the sub-halos, but this increase is observationally supressed by $(1+z)^{-1}$. This behaviour mirrors similar analyses of the ${\rm DM_{host}}$ contributions of host galaxies according to other cosmological simulations (\citealt{jaro20a,zhan20a}; {Kov\'{a}cs} et al., in prep.) 
However, we find that, on average, the accumulated DM due to traversing these sub-halos is lower that the $\sim90\,{\rm pc\;cm^{-3}}$ average excess DM which has been calculated for the purported likely galaxy-insersecting FRBs of \citet{conn22a}, lending weight to suggestions that FRBs could probe ionised matter within galaxy groups and clusters, and the feedback which informs it. Exploring the nature of this DM deficit may be an interesting avenue of research for future work. For example, \citet{simh20a} and \citet{lee21a} have already demonstrated that combining density reconstruction techniques with information about galaxies intersecting FRB sightlines may allow for better constraints on IGM DM contributions than the average values and variance offered by $\rm{DM_{cosmic}}$. \citet{ravi19a} and \citet{lee21a} both advocated complementing well-localised FRB observations with analysis of intervening halos along their sightlines, in order to constrain the fraction of baryons in the IGM, ${f_{\rm d}}$, and the distributions of baryons in CGM gas. In addition, by combining simulations of future dark matter halo catalogues, FRB redshift distributions, and free electron models for the IGM and galaxy halos, \citet{shir22a} have performed a detailed investigation of the joint constraining power of FRBs and dark matter halos. They conclude that cross-correlations between FRB DMs and galaxy cluster-sized halos ($M \sim10^{14}\,h^{-1}{\rm M}_{\odot}$) could potentially constrain multiple cosmological parameters including {$S_{8}(\equiv \sigma_8(\Omega_{\rm m}/0.3)^{0.5})$, $\Omega_{\rm b}$}, $h$, and $f_{\rm e}$; while cross-correlations with main sequence galaxy-sized halos ($M \sim10^{12}\,h^{-1}{\rm M}_{\odot}$) could inform models of galaxy formation and AGN feedback. 

Aside from simply offering a better understanding of the matter distribution of the Universe, learning more about cosmological LSS and the numbers, and natures, of collapsed structures along FRB sightlines may prove useful in hitherto unanticipated ways. To aid future work, we provide our results in various forms. For easy comparison to future analyses which will inevitably use other simulations as they are developed and refined, and for simple analysis of FRB observations, we provide average DM and sub-halo information derived from Figs.\ \ref{fig:LSSfracts}, \ref{fig:substats} and \ref{fig:subDMs} in Tables \ref{table:results}, \ref{table:resultsTraversed}, \ref{table:resultsImpact} and \ref{table:resultsDM}. For more in-depth analyses which may require them, we provide our best-fit parameters for our Eqs.\ (\ref{eq:fil_void_fit_function}) and (\ref{eq:halo_fit_function}) fits to our $\rm{DM^{TNG}_{halos}}$, $\rm{DM^{TNG}_{filaments}}$ and $\rm{DM^{TNG}_{voids}}$ distributions in Table \ref{table:results2}. We hope this information may prove useful in future analyses of cosmological ionised matter distributions, AGN feedback models, cosmological parameters, and FRBs themselves.

\begin{acknowledgements}
      {CRHW acknowledges invaluable help with TNG from Dylan Nelson, and from MPI and MPCDF staff including Markus Rampp, Thorsten Naab, Rudiger Pakmor, and members of the TRIDENT and YT project communities, including Britton Smith, Maan Hani, and John ZuHone. CRHW also thanks Stefan Hackstein, Xavier Prochaska, Sunil Simha and Mohit Bhardwaj for useful discussions. CRHW is a member of the Max Planck Lise Meitner group, and acknowledges support from the Max Planck Society. LGS is a Max Planck Lise Meitner group leader and acknowledges support from the Max Planck Society. YZM is supported by the National Research Foundation of South Africa under Grants No. 150580, No. 120385 and No. 120378, and the NITheCS program ``New Insights into Astrophysics and Cosmology with Theoretical Models confronting Observational Data''.  MCA acknowledges partial financial support from the Seal of Excellence @UNIPD 2020 program under the ACROGAL project. CBH is supported by NSF grant AAG-1911233, and NASA grants HST-AR-15800, HST-AR-16633, and HST-GO-16703. The authors thank the anonymous referee for valuable feedback during submission of this work. All data processing for this work was performed on the MPCDF high-power computing cluster, \texttt{RAVEN}\footnote{\url{https://www.mpcdf.mpg.de/services/supercomputing/raven}}.}
\end{acknowledgements}


\bibliographystyle{aa} 
\bibliography{references} 

\begin{table*}[t]
\begin{landscape}

\caption{Table of parameters used to fit Eqs.\ (\ref{eq:fil_void_fit_function}), (\ref{eq:halo_fit_function}) and (\ref{eq:hal_fil_void_fit_function2}) to the DM distributions shown in Fig.\ \ref{fig:LSSDMbreakdown} as a function of redshift.}\label{table:results2}
\centering
\resizebox{7in}{!}{\begin{tabular}{llllllllll}
\hline
\text{$z$} & \text{$\rm{a_{hal}}$} & \text{$\rm{l_{hal}}$} & \text{$\rm{s_{hal}}$} & \text{$\rm{a_{fil}}$} & \text{$\rm{l_{fil}}$} & \text{$\rm{s_{fil}}$} & \text{$\rm{a_{voi}}$} & \text{$\rm{l_{voi}}$} & \text{$\rm{s_{voi}}$}\\
\hline
0.10 & 6.12 & -2.66 & 4.69 & 0.51 & 11.76 & 47.38 & 0.26 & 8.07 & 4.89\\
0.20 & 13.15 & -8.22 & 10.72 & 0.46 & 29.34 & 94.33 & 0.25 & 17.59 & 8.76\\
0.30 & 54.14 & -41.75 & 44.62 & 0.40 & 47.94 & 146.43 & 0.22 & 27.51 & 13.16\\
0.40 & 191544.89 & -155041.91 & 155045.08 & 0.36 & 69.50 & 199.68 & 0.19 & 38.03 & 17.64\\
0.50 & 3651.94 & -2831.64 & 2835.08 & 0.32 & 91.92 & 255.21 & 0.18 & 49.10 & 22.03\\
0.70 & 226562.57 & -166369.30 & 166373.14 & 0.29 & 133.98 & 364.03 & 0.16 & 71.72 & 29.46\\
1.00 & 50183.27 & -34989.94 & 34994.16 & 0.28 & 225.75 & 498.55 & 0.18 & 112.03 & 34.93\\
1.50 & 1477835648.19 & -985206268.55 & 985206273.17 & 0.26 & 362.44 & 736.74 & 0.20 & 181.42 & 40.61\\
2.00 & 301737876.50 & -187118519.72 & 187118524.61 & 0.25 & 532.32 & 937.79 & 0.17 & 238.47 & 55.38\\
3.01 & 34.26 & -14.00 & 19.13 & 0.21 & 836.56 & 1338.24 & 0.20 & 356.55 & 57.30\\
4.01 & 434.53 & -234.63 & 239.93 & 0.19 & 1167.00 & 1678.88 & 0.07 & 322.65 & 180.45\\
5.00 & 28477.22 & -15415.56 & 15420.92 & 0.17 & 1522.72 & 1942.90 & 0.05 & 285.07 & 277.52\\
\end{tabular}}
\hfill
\bigskip
\bigskip

\end{landscape}

\caption{Table of LSS DM statistics (means, standard deviations, average fractions) for FRBs originating at given redshifts, from Fig.\ \ref{fig:LSSfracts}.}\label{table:results}

\centering

\resizebox{7in}{!}{\begin{tabular}{llllllll}

\hline
\noalign{\smallskip}
\text{$z$} & \text{$\rm{ \langle DM_{TNG}\rangle}$} & \text{$\rm{ \langle DM_{halos}\rangle}$} & \text{$\rm{ \langle DM_{filaments}\rangle}$} & \text{$\rm{ \langle DM_{voids}\rangle}$} & \text{$\langle f^{\rm DM}_{\rm halos}\rangle$} & \text{$\langle f^{\rm DM}_{\rm filaments}\rangle$} & \text{$\langle f^{\rm DM}_{\rm voids}\rangle$} \\
  & \text{$(\pm1\sigma)$} & \text{$(\pm1\sigma)$} & \text{$(\pm1\sigma)$} & \text{$(\pm1\sigma)$} & \text{$(\pm1\sigma)$} & \text{$(\pm1\sigma)$} & \text{$(\pm1\sigma)$} \\
\noalign{\smallskip}
\hline
0.00 & 0.00\,(0.00) & 0.00\,(0.00) & 0.00\,(0.00) & 0.00\,(0.00) & $-$ & $-$ & $-$ \\
0.10 & 92.59\,(63.30) & 13.64\,(47.60) & 65.82\,(29.20) & 13.13\,(1.34) & 0.15\,(0.13) & 0.71\,(0.11) & 0.14\,(0.06) \\
0.20 & 188.09\,(104.84) & 27.21\,(76.22) & 134.25\,(50.87) & 26.64\,(2.32) & 0.14\,(0.13) & 0.71\,(0.10) & 0.14\,(0.05) \\
0.30 & 288.67\,(135.56) & 41.64\,(107.54) & 206.68\,(66.43) & 40.99\,(2.99) & 0.14\,(0.21) & 0.72\,(0.10) & 0.14\,(0.05) \\
0.40 & 393.05\,(159.92) & 55.95\,(148.31) & 282.40\,(78.74) & 56.01\,(3.52) & 0.14\,(0.27) & 0.72\,(0.10) & 0.14\,(0.04) \\
0.50 & 500.30\,(180.34) & 69.29\,(159.22) & 360.85\,(89.31) & 71.47\,(3.95) & 0.14\,(0.22) & 0.72\,(0.09) & 0.14\,(0.04) \\
0.70 & 708.06\,(216.37) & 94.03\,(176.49) & 513.76\,(112.50) & 101.58\,(4.87) & 0.13\,(0.17) & 0.73\,(0.09) & 0.14\,(0.03) \\
1.00 & 1020.60\,(282.94) & 130.39\,(220.56) & 744.97\,(150.57) & 147.52\,(6.36) & 0.13\,(0.14) & 0.73\,(0.08) & 0.14\,(0.03) \\
1.50 & 1525.75\,(368.14) & 183.58\,(307.68) & 1125.31\,(203.45) & 222.93\,(8.37) & 0.12\,(0.13) & 0.74\,(0.07) & 0.15\,(0.03) \\
2.00 & 2011.22\,(420.97) & 227.92\,(375.51) & 1499.37\,(241.24) & 294.64\,(9.44) & 0.11\,(0.13) & 0.75\,(0.07) & 0.15\,(0.03) \\
3.0{\color{ForestGreen}1} & 2889.42\,(484.86) & 285.98\,(480.56) & 2205.80\,(293.81) & 414.93\,(11.31) & 0.10\,(0.12) & 0.76\,(0.06) & 0.14\,(0.02) \\
4.0{\color{ForestGreen}1} & 3674.05\,(515.59) & 314.90\,(497.89) & 2876.14\,(323.86) & 503.61\,(13.29) & 0.09\,(0.10) & 0.78\,(0.05) & 0.14\,(0.02) \\
5.00 & 4362.14\,(527.60) & 327.61\,(505.21) & 3494.53\,(339.35) & 563.02\,(14.93) & 0.08\,(0.09) & 0.80\,(0.05) & 0.13\,(0.02) \\

\end{tabular}}
\hfill
\end{table*}

\newpage

\begin{table}[t]
\caption{Table containing the number of unique sub-halos which are traversed by all line of sight segments generated for a given snapshot at a given redshift. A superscript `total' denotes the total number of sub-halos traversed by all segments for the snapshot. A superscript $[X,Y]$ denotes the number of these sub-halos which fall within a mass bin $M_{\rm sub}=[10^X,10^Y]\,h^{-1}{\rm M_{\odot}}$. These data are further visualised in Fig.\ \ref{fig:DMvIF}.}\label{table:segmentsubhalos}

\centering

\resizebox{3.5in}{!}{\begin{tabular}{llllll}

\hline
\noalign{\smallskip}
\text{$z$} & \text{${N^{\rm total}_{\rm sub}}$} & \text{${ N^{ [8,10]}_{\rm sub}}$} & \text{${N^{ [10,12]}_{\rm sub}}$} & \text{${N^{ [12,14]}_{\rm sub}}$} & \text{${N^{ [14,16]}_{\rm sub}}$}\\
\noalign{\smallskip}
\hline
0.00 & 7809 & 1546 & 4618 & 1514 & 131 \\
0.10 & 7836 & 1646 & 4552 & 1546 & 92 \\
0.20 & 8285 & 1801 & 4899 & 1469 & 116 \\
0.30 & 8562 & 1864 & 5048 & 1569 & 81 \\
0.40 & 8829 & 1975 & 5285 & 1510 & 59 \\
0.50 & 9292 & 2199 & 5485 & 1558 & 50 \\
0.70 & 9797 & 2634 & 5645 & 1478 & 40 \\
1.00 & 10474 & 3350 & 5864 & 1243 & 17 \\
1.50 & 12040 & 4738 & 6354 & 945 & 3 \\
2.00 & 13397 & 6279 & 6495 & 623 & 0 \\
3.01 & 14454 & 8793 & 5427 & 234 & 0 \\
4.01 & 14363 & 10617 & 3698 & 48 & 0 \\
5.00 & 14369 & 11901 & 2460 & 8 & 0 \\

\end{tabular}}
\hfill

\bigskip
\bigskip

\caption{Table of statistics (mean, standard deviation) describing the average number of sub-halos \textit{which are traversed by FRBs originating at given redshifts}. A superscript `total' denotes a statistic obtained by using all sub-halos we consider traversed by our segments. A superscript $[X,Y]$ denotes a statistic obtained by using only traversed sub-halos within a mass bin $M_{\rm sub}=[10^X,10^Y]\,h^{-1}{\rm M_{\odot}}$. These values are calculated from sightlines created according to Eq.\ (\ref{eq:discrete_subhalo}), and are shown in Fig.\ \ref{fig:substats} (left).}\label{table:resultsTraversed}

\centering

\resizebox{4.0in}{!}{\begin{tabular}{llllll}

\hline
\noalign{\smallskip}
\text{$z$} & \text{${ \langle N^{\rm total}_{sub}\rangle}$} & \text{${ \langle N^{ [8,10]}_{\rm sub}\rangle}$} & \text{${ \langle N^{ [10,12]}_{\rm sub}\rangle}$} & \text{${ \langle N^{ [12,14]}_{\rm sub}\rangle}$} & \text{${ \langle N^{ [14,16]}_{\rm sub}\rangle}$}\\
  & \text{$(\pm 1\sigma)$} & \text{$(\pm 1\sigma)$} & \text{$(\pm 1\sigma)$} & \text{$(\pm 1\sigma)$} & \text{$(\pm 1\sigma)$}\\
\noalign{\smallskip}
\hline
0.00 & 0.00\,(0.00) & 0.00\,(0.00) & 0.00\,(0.00) & 0.00\,(0.00) & 0.000\,(0.000) \\
0.10 & 0.15\,(0.09) & 0.03\,(0.04) & 0.09\,(0.07) & 0.03\,(0.04) & 0.002\,(0.010) \\
0.20 & 0.31\,(0.16) & 0.06\,(0.07) & 0.18\,(0.12) & 0.06\,(0.07) & 0.004\,(0.017) \\
0.30 & 0.48\,(0.21) & 0.10\,(0.09) & 0.28\,(0.16) & 0.09\,(0.09) & 0.006\,(0.023) \\
0.40 & 0.65\,(0.25) & 0.14\,(0.11) & 0.38\,(0.19) & 0.12\,(0.11) & 0.007\,(0.026) \\
0.50 & 0.83\,(0.29) & 0.18\,(0.13) & 0.49\,(0.21) & 0.15\,(0.12) & 0.009\,(0.028) \\
0.70 & 1.21\,(0.38) & 0.27\,(0.17) & 0.71\,(0.28) & 0.21\,(0.16) & 0.010\,(0.032) \\
1.00 & 1.80\,(0.54) & 0.45\,(0.27) & 1.05\,(0.39) & 0.29\,(0.21) & 0.011\,(0.035) \\
1.50 & 2.89\,(0.87) & 0.84\,(0.47) & 1.64\,(0.63) & 0.40\,(0.29) & 0.012\,(0.040) \\
2.00 & 4.12\,(1.20) & 1.36\,(0.68) & 2.27\,(0.87) & 0.48\,(0.37) & 0.012\,(0.040) \\
3.01 & 6.82\,(1.91) & 2.82\,(1.23) & 3.44\,(1.35) & 0.56\,(0.46) & 0.012\,(0.040) \\
4.01 & 9.66\,(2.69) & 4.74\,(1.86) & 4.32\,(1.71) & 0.58\,(0.49) & 0.012\,(0.040) \\
5.00 & 12.43\,(3.27) & 6.93\,(2.41) & 4.90\,(1.88) & 0.59\,(0.50) & 0.012\,(0.040) \\

\end{tabular}}
\hfill

\end{table}

\newpage

\begin{landscape}
\begin{table}[t]

\caption{Table of observed impact parameter statistics (median, interquartile range, IQR, and interpercentile range, IPR, between the 10th and 90th percentiles) from Fig.\ \ref{fig:substats} (right). All values are in units of $h^{-1}\,{\rm ckpc}$. A superscript `total' denotes a statistic obtained by using sub-halos of any mass which we consider traversed by our segments. A superscript $[X,Y]$ denotes a statistic obtained by using only traversed sub-halos within a mass bin $M_{\rm sub}=[10^X,10^Y]\,h^{-1}{\rm M_{\odot}}$. Statistics for a particular mass bin and redshift are only provided when the total number of sub-halos traversed by segments in the snapshot for this mass bin is $\ge40$ (see Table \ref{table:segmentsubhalos}).}\label{table:resultsImpact}

\centering

\resizebox{9.5in}{!}{\begin{tabular}{llllllllllllllll}

\hline
\noalign{\smallskip}
\text{$z$} & \text{${\rm m}({b}^{\rm total}_{\perp,{\rm sub}})$} & \text{${\rm IQR^{total}}$} & \text{${\rm IPR^{total}}$} & \text{${\rm m}({b}^{\rm [8,10]}_{\perp,{\rm sub}})$} & \text{${\rm IQR^{[8,10]}}$} & \text{${\rm IPR^{[8,10]}}$} & \text{${\rm m}({b}^{\rm [10,12]}_{\perp,{\rm sub}})$} & \text{${\rm IQR^{[10,12]}}$} & \text{${\rm IPR^{[10,12]}}$} & \text{${\rm m}({b}^{\rm [12,14]}_{\perp,{\rm sub}})$} & \text{${\rm IQR^{[12,14]}}$} & \text{${\rm IPR^{[12,14]}}$} & \text{${\rm m}({b}^{\rm [14,16]}_{\perp,{\rm sub}})$} & \text{${\rm IQR^{[14,16]}}$} & \text{${\rm IPR^{[14,16]}}$}\\
\noalign{\smallskip}
\hline
0.0 & 72.33 & $[41.33,131.86]$ & $[23.06,303.08]$ & 36.14 & $[22.18,52.76]$ & $[13.09,70.21]$ & 69.32 & $[44.44,100.71]$ & $[26.44,138.87]$ & 263.18 & $[171.73,443.94]$ & $[101.99,641.34]$ & 949.96 & $[619.41,1388.02]$ & $[381.71,1729.83]$ \\
0.1 & 72.13 & $[40.50,134.97]$ & $[22.83,305.86]$ & 34.99 & $[21.44,53.04]$ & $[13.22,72.46]$ & 70.08 & $[44.81,102.02]$ & $[27.62,145.16]$ & 281.00 & $[182.70,455.66]$ & $[108.83,647.82]$ & 990.77 & $[656.89,1354.89]$ & $[462.70,1782.73]$ \\
0.2 & 69.61 & $[40.42,128.79]$ & $[22.96,293.68]$ & 36.40 & $[22.24,53.34]$ & $[13.13,70.40]$ & 69.35 & $[44.74,102.22]$ & $[27.52,142.67]$ & 282.65 & $[177.09,454.83]$ & $[110.72,701.42]$ & 1068.78 & $[808.41,1348.18]$ & $[556.35,2020.01]$ \\
0.3 & 70.83 & $[40.00,134.42]$ & $[22.67,281.32]$ & 35.87 & $[22.67,51.93]$ & $[14.01,70.69]$ & 70.30 & $[45.18,106.27]$ & $[26.28,149.62]$ & 274.46 & $[177.85,441.57]$ & $[110.38,678.69]$ & 1093.50 & $[732.67,1365.93]$ & $[402.98,1799.59]$ \\
0.4 & 70.00 & $[39.88,128.27]$ & $[22.60,268.70]$ & 35.13 & $[21.66,50.99]$ & $[12.97,69.07]$ & 71.76 & $[46.36,105.07]$ & $[27.83,149.92]$ & 280.06 & $[184.85,438.88]$ & $[112.76,663.54]$ & 902.54 & $[639.83,1377.15]$ & $[268.27,1915.76]$ \\
0.5 & 69.31 & $[39.78,128.97]$ & $[22.57,273.46]$ & 35.18 & $[22.35,51.86]$ & $[13.30,70.34]$ & 71.43 & $[45.96,106.64]$ & $[28.00,150.60]$ & 291.67 & $[197.19,454.05]$ & $[118.14,704.00]$ & 1219.48 & $[1056.59,1564.19]$ & $[782.31,1829.96]$ \\
0.7 & 67.02 & $[38.47,123.44]$ & $[22.11,247.20]$ & 35.46 & $[22.49,51.54]$ & $[13.61,68.77]$ & 73.21 & $[47.19,110.19]$ & $[29.07,153.92]$ & 278.30 & $[190.28,429.88]$ & $[109.09,633.66]$ & 1253.84 & $[776.75,1578.67]$ & $[597.97,2065.59]$ \\
1.0 & 61.67 & $[35.38,110.87]$ & $[20.25,211.58]$ & 34.65 & $[21.91,51.09]$ & $[13.50,68.85]$ & 72.03 & $[45.68,110.15]$ & $[28.42,156.86]$ & 271.98 & $[180.98,412.11]$ & $[110.69,595.14]$ & $-$ & $-$ & $-$ \\
1.5 & 56.83 & $[32.65,98.69]$ & $[18.58,178.76]$ & 34.87 & $[21.75,50.93]$ & $[12.89,68.31]$ & 74.60 & $[46.44,112.56]$ & $[28.30,162.99]$ & 269.64 & $[188.29,398.96]$ & $[117.18,576.70]$ & $-$ & $-$ & $-$ \\
2.0 & 51.74 & $[30.01,87.05]$ & $[17.56,146.71]$ & 34.36 & $[21.90,50.88]$ & $[12.88,68.78]$ & 73.63 & $[47.08,111.87]$ & $[28.58,160.56]$ & 238.81 & $[162.85,365.65]$ & $[104.35,519.97]$ & $-$ & $-$ & $-$ \\
3.0 & 44.98 & $[26.92,74.08]$ & $[15.62,118.51]$ & 34.05 & $[21.29,50.01]$ & $[12.80,68.23]$ & 75.00 & $[47.97,112.80]$ & $[29.50,158.13]$ & 251.19 & $[175.08,333.56]$ & $[115.35,459.17]$ & $-$ & $-$ & $-$ \\
4.0 & 40.54 & $[24.88,63.36]$ & $[14.73,93.26]$ & 34.60 & $[21.88,50.65]$ & $[13.29,68.74]$ & 70.63 & $[45.95,104.66]$ & $[27.86,149.57]$ & 251.66 & $[200.73,330.89]$ & $[144.87,485.22]$ & $-$ & $-$ & $-$ \\
5.0 & 41.41 & $[26.39,60.69]$ & $[15.99,85.09]$ & 37.94 & $[24.52,54.03]$ & $[15.07,71.96]$ & 70.58 & $[46.32,102.48]$ & $[28.32,142.36]$ & $-$ & $-$ & $-$ & $-$ & $-$ & $-$ \\

\end{tabular}}
\hfill
\end{table}

\bigskip

\begin{table}[t]

\caption{Table of observed DM statistics (median, interquartile range, IQR, and interpercentile range, IPR, between the 10th and 90th percentiles) from Fig.\ \ref{fig:subDMs} (right). All values are in units of ${\rm pc\;cm^{-3}}$. A superscript `total' denotes a statistic obtained by using sub-halos of any mass which we consider traversed by our segments. A superscript $[X,Y]$ denotes a statistic obtained by using only traversed sub-halos within a mass bin $M_{\rm sub}=[10^X,10^Y]\,h^{-1}{\rm M_{\odot}}$. Statistics for a particular mass bin and redshift are only provided when the total number of sub-halos traversed by segments in the snapshot for this mass bin is $\ge40$ (see Table \ref{table:segmentsubhalos}).}\label{table:resultsDM}

\centering

\resizebox{9.5in}{!}{\begin{tabular}{llllllllllllllll}

\hline
\noalign{\smallskip}
\text{$z$} & \text{${\rm m}({\rm DM}^{\rm total}_{{\rm sub}})$} & \text{${\rm IQR^{total}}$} & \text{${\rm IPR^{total}}$} & \text{${\rm m}({\rm DM}^{\rm [8,10]}_{{\rm sub}})$} & \text{${\rm IQR^{[8,10]}}$} & \text{${\rm IPR^{[8,10]}}$} & \text{${\rm m}({\rm DM}^{\rm [10,12]}_{{\rm sub}})$} & \text{${\rm IQR^{[10,12]}}$} & \text{${\rm IPR^{[10,12]}}$} & \text{${\rm m}({\rm DM}^{\rm [12,14]}_{{\rm sub}})$} & \text{${\rm IQR^{[12,14]}}$} & \text{${\rm IPR^{[12,14]}}$} & \text{${\rm m}({\rm DM}^{\rm [14,16]}_{{\rm sub}})$} & \text{${\rm IQR^{[14,16]}}$} & \text{${\rm IPR^{[14,16]}}$}\\
\noalign{\smallskip}
\hline
0.0 & 1.00 & $[0.28,4.46]$ & $[0.12,17.73]$ & 0.22 & $[0.10,0.50]$ & $[0.06,1.09]$ & 1.03 & $[0.34,3.17]$ & $[0.15,9.29]$ & 7.47 & $[2.03,24.52]$ & $[0.56,69.09]$ & 36.12 & $[4.20,116.17]$ & $[1.61,323.59]$ \\
0.1 & 1.07 & $[0.31,4.65]$ & $[0.13,17.32]$ & 0.26 & $[0.12,0.60]$ & $[0.07,1.26]$ & 1.12 & $[0.37,3.49]$ & $[0.16,10.57]$ & 7.63 & $[2.07,22.68]$ & $[0.61,51.86]$ & 29.63 & $[5.79,131.98]$ & $[1.56,376.27]$ \\
0.2 & 1.18 & $[0.33,4.87]$ & $[0.14,18.27]$ & 0.30 & $[0.14,0.72]$ & $[0.08,1.62]$ & 1.23 & $[0.40,3.89]$ & $[0.17,10.79]$ & 8.06 & $[2.69,24.01]$ & $[0.80,57.09]$ & 24.39 & $[5.48,106.01]$ & $[1.58,258.36]$ \\
0.3 & 1.32 & $[0.39,5.15]$ & $[0.16,18.91]$ & 0.35 & $[0.16,0.86]$ & $[0.09,2.06]$ & 1.39 & $[0.47,4.22]$ & $[0.20,12.22]$ & 8.10 & $[2.31,24.80]$ & $[0.74,64.56]$ & 20.58 & $[2.54,111.11]$ & $[0.70,467.00]$ \\
0.4 & 1.27 & $[0.40,5.05]$ & $[0.18,18.39]$ & 0.43 & $[0.19,1.00]$ & $[0.11,2.22]$ & 1.31 & $[0.45,4.22]$ & $[0.20,12.24]$ & 9.10 & $[2.64,26.84]$ & $[0.84,72.00]$ & 77.47 & $[13.20,440.62]$ & $[5.13,743.17]$ \\
0.5 & 1.37 & $[0.42,5.31]$ & $[0.19,18.72]$ & 0.48 & $[0.21,1.09]$ & $[0.11,2.46]$ & 1.48 & $[0.49,4.72]$ & $[0.22,12.91]$ & 8.68 & $[2.41,28.15]$ & $[0.85,75.98]$ & 17.09 & $[1.82,52.93]$ & $[0.52,95.31]$ \\
0.7 & 1.50 & $[0.48,5.80]$ & $[0.21,20.72]$ & 0.62 & $[0.26,1.44]$ & $[0.15,3.43]$ & 1.67 & $[0.55,5.50]$ & $[0.24,15.86]$ & 10.98 & $[3.11,32.98]$ & $[0.95,74.10]$ & 41.78 & $[11.71,95.03]$ & $[0.69,372.61]$ \\
1.0 & 1.63 & $[0.55,6.21]$ & $[0.25,21.49]$ & 0.75 & $[0.34,1.85]$ & $[0.18,4.60]$ & 2.00 & $[0.68,6.66]$ & $[0.29,19.17]$ & 12.77 & $[3.53,39.48]$ & $[1.00,84.61]$ & $-$& $-$ & $-$ \\
1.5 & 1.90 & $[0.66,6.38]$ & $[0.31,20.91]$ & 1.10 & $[0.47,2.82]$ & $[0.25,6.85]$ & 2.53 & $[0.81,8.05]$ & $[0.36,23.12]$ & 14.02 & $[3.82,41.15]$ & $[1.10,86.18]$ & $-$ & $-$ & $-$ \\
2.0 & 2.03 & $[0.75,6.47]$ & $[0.36,19.98]$ & 1.41 & $[0.60,3.42]$ & $[0.31,8.54]$ & 2.84 & $[0.96,9.26]$ & $[0.40,25.61]$ & 18.94 & $[5.27,51.31]$ & $[1.63,96.16]$ & $-$ & $-$ & $-$ \\
3.0 & 2.42 & $[0.97,6.70]$ & $[0.47,18.03]$ & 1.90 & $[0.84,4.53]$ & $[0.43,10.40]$ & 3.70 & $[1.27,10.96]$ & $[0.55,27.25]$ & 19.93 & $[4.91,57.28]$ & $[1.62,110.65]$ & $-$ & $-$ & $-$ \\
4.0 & 2.70 & $[1.11,6.85]$ & $[0.52,15.01]$ & 2.36 & $[1.02,5.34]$ & $[0.48,11.08]$ & 4.62 & $[1.53,13.74]$ & $[0.64,27.67]$ & 21.92 & $[6.99,51.77]$ & $[1.86,69.51]$ & $-$ & $-$ & $-$ \\
5.0 & 2.40 & $[0.98,5.44]$ & $[0.45,10.36]$ & 2.12 & $[0.90,4.89]$ & $[0.43,8.69]$ & 4.64 & $[1.67,10.85]$ & $[0.69,18.80]$ & $-$ & $-$ & $-$ & $-$ & $-$ &  $-$ \\

\end{tabular}}
\hfill
\end{table}

\end{landscape}

\end{document}